\documentclass[usenatbib]{mn2e}

\usepackage{graphicx}
\usepackage{amssymb}

\title[Excluding interlopers from asteroid families]{An automatic approach to exclude interlopers from asteroid families}

\author[Radovi\'c et al.]{Viktor Radovi\' c,$^{1}$\thanks{E-mail: rviktor@matf.bg.ac.rs (VR)}
Bojan Novakovi\' c,$^{1}$
Valerio Carruba$^{2}$
\& Du\v san Mar\v ceta$^{1}$
\\
$^{1}$Department of Astronomy, Faculty of Mathematics, University of Belgrade, Studentski trg 16, 11000 Belgrade, Serbia\\
$^{2}$UNESP, Univ. Estadual Paulista, Grupo de dinamica Orbital e Planetologia, Guaratinguetá, SP, 12516-410, Brazil\\
}

\begin{document}

\date{Accepted . Received ; in original form }

\pagerange{\pageref{firstpage}--\pageref{lastpage}} \pubyear{2017}

\maketitle

\label{firstpage}

\begin{abstract}
Asteroid families are valuable source of information to many asteroid-related researches, assuming a reliable list of their members could be obtained.  However, as the number of known asteroids increases fast it becomes more and more difficult to obtain robust list of members of an asteroid family.
Here we are proposing a new approach to deal with the problem, based on the well known Hierarchical Clustering Method (HCM). An additional step in the whole procedure is introduced in order to reduce a so-called chaining effect. The main idea is to prevent chaining through an
already identified interloper. We show that in this way a number of potential interlopers among family members is significantly reduced. 
Moreover, we developed an automatic on-line based portal to apply this procedure, i.e to generate a list of family members as well as a list of 
potential interlopers. The Asteroid Families Portal (AFP) is freely available to all interested researchers.
\end{abstract}

\begin{keywords}
asteroids: general
\end{keywords}

\section{Introduction}
\label{s:intro}

Asteroid families are formed when a collision breaks apart a parent body into numerous 
smaller fragments \citep{zappala1984}. As such, they are very important for almost any 
asteroid-related research. For example, families may provide us a clue about collisional 
history of the main asteroid belt \citep[see e.g.][and references therein]{cibulkova2014}, 
asteroid interiors \citep{cellino2002}, as well as space weathering effects that alter 
asteroids colour with time \citep{nes2005}. Moreover, they may even tell us about the origins 
of some near-Earth asteroids \citep[e.g.][]{walsh2013,bottke2015,nov2017}, main-belt comets \citep{nov2014,hsieh2015}, 
or help us put constraints on some past phenomena like the Late Heavy Bombardment \citep{broz2013}.

The impact ejects fragments away from the parent body at velocities similar to the parent's body 
escape speed, that is typically not more than a few tens of meters per second. In the main asteroid 
belt these velocities are much slower than the orbital ones, which are about $15 - 20\ km/s$. 
Therefore, the impact-produced objects initially keep orbits similar to the orbit of their parent body. 
This fact allows identification of the so-called dynamical asteroid families, i.e. groups of asteroids 
identified based on similarity of their orbital parameters. Still, from the orbital similarity point 
of view, it is important to distinguish young asteroid families from older ones. 

Very young families ($t_{age} < 1 m.y. $) may be still recognized in the space of instantaneous (osculating) orbital elements, because different perturbations have not yet had enough time to disperse orbits of their members \citep{nes2006}. On the other hand, older families are usually identified using the so-called proper orbital elements which are nearly constant over time \citep{knezevic2000}.
 
So far a large number of asteroid families have been discovered across the main asteroid belt 
\citep[e.g.][]{zappala1990, nes2005, novakovic2011, masiero2013, milani2014,nes2015,milani2017}. These families offer 
a wide range of possibilities for further studies, yet an essential prerequisite for all these studies
is a reliably established family membership. However, as nowadays the number of known asteroids 
increases fast, it becomes more and more difficult to obtain a robust list of members of an 
asteroid family.

For the purpose of family identification the Hierarchical Clustering Method (HCM), proposed by \citet{zappala1990}, 
is most widely used.\footnote{Besides the HCM, the Wavelet Analysis Method (WAM) 
was also successfully applied to identify asteroid families \citep{bendjoya1991}.}
It connects all objects whose mutual distances in the three-dimensional proper-element space
are below a threshold value.

The HCM, however, obviously has some limitations in distinguishing between real family members and nearby background asteroids.\footnote{This problem is not due the HCM itself, and is not limited to this method only. For instance, the WAM suffers from the same problem \citep{Zappala1995}.}  This gives a rise to the well known
problem of presence of interlopers among the members of an asteroid family \citep{migliorini1995}.

A possible way to deal with this problem is to use additional information to discriminate
real family members from interlopers. Asteroids that belong to the same family generally 
have similar mineral composition \citep{cellino2002}. Thus, data about their spectra/colours or 
albedos may help to determine membership of an asteroid family more reliably. 

These additional information may be used in different ways. One possible method to exploit 
available physical data is to apply the HCM in \textit{extended} space, i.e. in the space that
also includes physical data in addition to the three proper elements. \citet{parker2008} applied
the HCM in 4-dimensional space, using the Sloan Digital Sky Survey \citep[SDSS;][]{york2000}
colours as the fourth dimension. \citet{carruba2013} extend this approach to the 5th dimension 
using albedos as an additional dimension, significantly reducing the percentage of known interlopers with respect to other methods.

Another possible procedure is to first separate the main belt asteroids into two
populations (typically representing $C-$ and $S-$type objects) according to their 
colour and albedo values. The HCM is then applied to each of these populations
separately \citep{masiero2013}. In this way Masiero et al. managed to identify several
new families.

These methods, despite being in general very efficient, have a serious limitation, 
that they can only be applied to a reduced set of main
belt asteroids for which the colours and albedos are obtained. 
Despite significant increase of available physical data in the recent years, 
the number of asteroids for which these data are at our disposal is still several times smaller
than the number of objects for which proper elements have been computed. As explained by \citet{milani2014}, the dynamical parameters, in this case the proper
elements, have a larger information content than the physical observations, because the latter are available either for significantly smaller catalogues, or with lower relative accuracy.

For these reasons some authors adopted a bit different approach, so that
the so-called dynamical families are first obtained in the space of proper elements,
and available physical data is used only posteriori to distinguish possible
overlapping families or to identify interlopers among family members \citep[see e.g.][]{novakovic2011,milani2014,milani2016}. 

All the above described methodologies are useful to some extent. Still, they have
some limitations imposed by the HCM itself. A well-known drawback of the HCM based on 
the single linkage rule is the so-called \textit{chaining} phenomenon; that is, 
first concentrations tend to incorporate nearby groups, forming as a result a kind of 
\textit{chain} which may consists of non-family members \citep{zappala1994}.

In this paper we extend previous works in two directions: 
\begin{itemize}
\item First, we are introducing an additional step in the whole procedure, aiming
mainly to reduce the chaining effect. The main idea is that if an interloper is
identified as a family member, it may cause more interlopers to be added
due to the chaining effect. In this manner we managed to further reduce the number of 
potential interlopers among family members, by preventing chaining through an
already identified interloper.
\item Second, we are presenting here an automatic, free, on-line based tool
to apply our procedure. It allows to generate a list of family members 
as well as a list of potential interlopers along with criterion for their rejection.
This is important because despite numerous papers dealing with family members 
identification and removal of potential interlopers, there is still a 
lack of publicly available information on these. 
\end{itemize}

\section[]{Method}
\label{s:method}

\begin{figure}
	 \includegraphics[width=84mm]{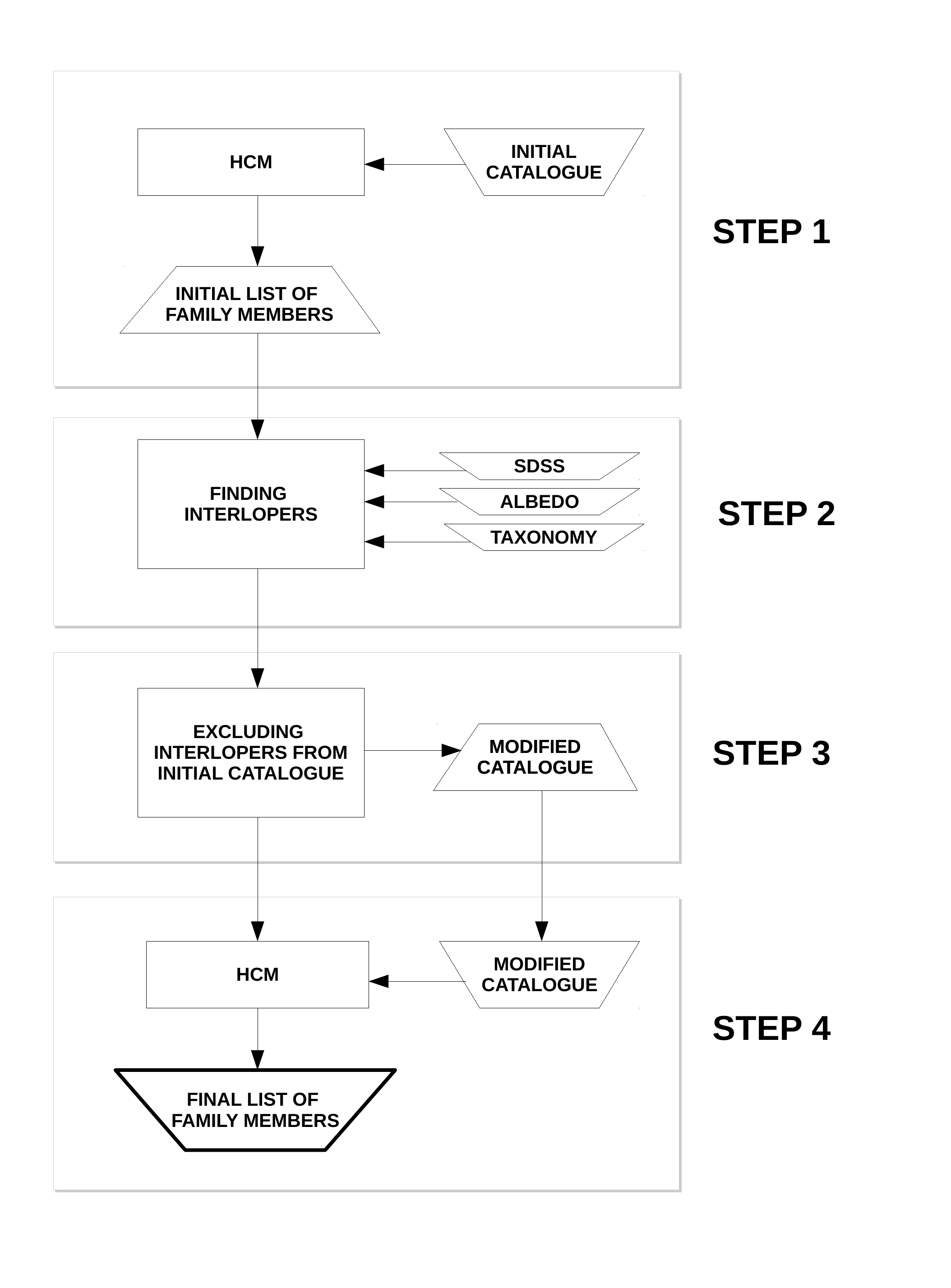}
 \caption{The chart showing the algorithm of our methodology to remove interlopers from asteroid families. }
 \label{f:chart}
\end{figure}

The approach that we are using here is essentially very similar to methods that first
identify dynamical families using only the proper orbital elements, and
then apply physical data to further refine family membership. The only
exception in this respect is an additional stage.
The whole procedure could be divided in four main stpdf (Fig.~\ref{f:chart}):
\begin{itemize}
\item In the first step, the HCM analysis is performed using the whole (initial) catalogue of 
proper elements in order to obtain a preliminary list of family members.
\item In the second step, physical and spectral properties are used in order to identify interlopers 
among asteroids initially linked to a family.
\item In the third step, objects identified as interlopers in the second step, are excluded from 
the initial catalogue of proper elements, producing a modified catalogue
\item Finally, in the fourth step, the HCM analysis is performed again, but this time using the reduced (modified) 
catalogue of proper elements, produced in the third step.
\end{itemize} 

\subsection{STEP 1: Obtaining a preliminary membership}
\label{ss:step1}

To obtain a preliminary list of family members we apply the HCM in the space of three
proper orbital elements\footnote{The catalogue of synthetic proper elements is available at
AstDys web page: http://hamilton.dm.unipi.it}: semi-major axis $a_p$, eccentricity $e_p$, 
and sine of inclination $\sin (i_p)$. The HCM identifies an asteroid as a part of 
family if its distance $d$ from the closest neighbour is smaller than an adopted cut-off 
distance $d_{cut}$. The common definition of this metric is \citep{zappala1990}:
\begin{equation}
d = na_p\sqrt{C_a (\delta a_p/a_p)^2 + C_e (\delta e_p)^2 + C_i (\delta \sin(i_p))^2}
\end{equation}
where $na_p$ is the heliocentric orbital velocity of an asteroid on a circular orbit having the semi-major axes $a_p$, $\delta a_p = a_{p_1} - a_{p_2}$, $\delta e_p = e_{p_1} - e_{p_2}$ and $\delta \sin(i_p) = \sin(i_{p_1}) - \sin(i_{p_2})$, where the indexes (1) and (2) denote the two bodies whose mutual distance is calculated.  The distance $d$ is expressed in meters per second.

The above metric is derived based on the well-known relationship between the components of ejection velocities of the fragments, and the resulting differences in their orbital elements (Gaussian equations).
For the constants, the most frequently used values are: $C_a = \frac{5}{4}$, $C_e = 2$ and $C_i = 2$ \citep{zappala1994}, but other values yield the similar results \citep{nes2005}. 

Here we applied the HCM around a selected central asteroid. In this case the volume of the proper elements space is not fixed a priory, but it grows around the central object as the velocity 
cut-off distance increases.
In order to define a preliminary list of family members we need to adopt an appropriate value of
cut-off distance $d_{cut}$. Generally, this is the distance in the space of proper elements that best describes a family.  However, for a preliminary definition of the family our aim is 
slightly different. As we want to identify as many potential interlopers as possible, we 
used the largest reasonable value of $d_{cut}$, rather than the most appropriate one.

Thus, in the first step we derive the cut-off value as follows. Starting from the lowest 
value of $10~ms^{-1}$, we increased $d_{cut}$ by fixed step of $5~ms^{-1}$, until the family merges 
with a local background population of asteroids. For the threshold value we adopt the one two stpdf 
($10~ms^{-1}$) below the distance at which the family merges with the background. 

As mentioned above, the cut-off value obtained in this way is often too large to define a nominal family.
In some cases this approach caused two (or even more) nearby families to merge into single group 
(see Sections~\ref{ss:other_fam} and \ref{ss:special_cases} for discussion on these cases).
Still, we use it in order to get as many family members (and consequently also interlopers) as possible. 
 
\subsection{STEP 2: Interlopers identification}
\label{ss:interlopers_identification}

Once a preliminary list of family members is obtained, the next step is the identification of potential interlopers among them. In this respect, let us recall the well known fact that members of a collisional 
family typically share similar physical and spectral characteristics \citep[][]{Bus1999, Lazzaro1999, ivezic2002, cellino2002}. Therefore, different spectral and photometric data could 
be used to complement the results of the HCM analysis, in order to obtain a list of potential interlopers. 

Available observational data is often good enough only to distinguish between $C$ and $S$ classes of asteroids, while other taxonomy classes could not be separated reliably. An obvious exception are spectral data, which are however available for a very limited number of asteroids. The latter may also be true for an advance classification algorithm developed by \citet{DeMeoCarry2013}, 
based on the SDSS data.
Still, to stay on the safe side, we decided to resolve here only two aforementioned broad 
asteroid classes. This analysis is 
performed using SDSS colours, geometric albedos from different surveys, and available 
spectroscopic data. Then, if it is found that a family is mostly composed of $C$-type asteroids, 
all potential members that belong to $S$-type are defined as interlopers, and vice-versa.

\subsubsection{The Sloan Digital Sky Survey data}

The SDSS survey used 5-colour photometric system $u$, $g$, $r$, $i$ and $z$, with central wavelengths of $0.3543$, $0.4770$, $0.6231$, $0.7625$, $0.9134$~$\mu$m, respectively. The fourth release of the SDSS Moving Object Catalogue (SDSS MOC) contains photometric data for 471,569 moving objects observed 
up to March, 2007. Among this data, 220,101 entries were successfully linked to 104,449 unique moving objects. 

It was shown by \citet{ivezic2002} that SDSS photometry is consistent with available spectra of asteroids,
meaning that the colours provided by the SDSS could be used to separate at least broad taxonomic 
classes, such as $C$ and $S$. This was also confirmed by \citet{nes2005} and \citet{parker2008} who 
used the third and fourth release of the SDSS MOC, respectively. In both cases authors used a 
principal component analysis approach to link SDSS colours to spectral types. The main difference between the method utilized by \citet{nes2005} and \citet{parker2008} is that latter authors excluded $u$-band from their analysis, due to the large noise in $u$-band presented in the forth version of SDSS MOC.

Moreover, \citet{carvano2010} defined a new classification algorithm
based on the SDSS colours that can be used to assign taxonomic 
classes to SDSS observations. This scheme is compatible 
with the Bus taxonomy and allows finer distinction between
taxonomic classes than methodology used by \citet{nes2005} or \citet{parker2008}.

For our purpose here, we chose to use the fourth release of the SDSS MOC
to distinguish between $C$ and $S$ taxonomic complexes. Thus, we adopt
\citet{parker2008} approach excluding $u$-band from our analysis, and using 
$a^{*}$ colour defined by \citet{ivezic2001} as:
\begin{equation}
a^* =   0.89 (g - r) + 0.45 (r - i) - 0.57
\end{equation}

It is known that asteroids show bimodal distribution in $a^*$, where $C$-type objects 
are characterized with $a^*< 0$, while $S$-type objects typically have $a^* > 0$.

\begin{figure}
\includegraphics[width=84mm]{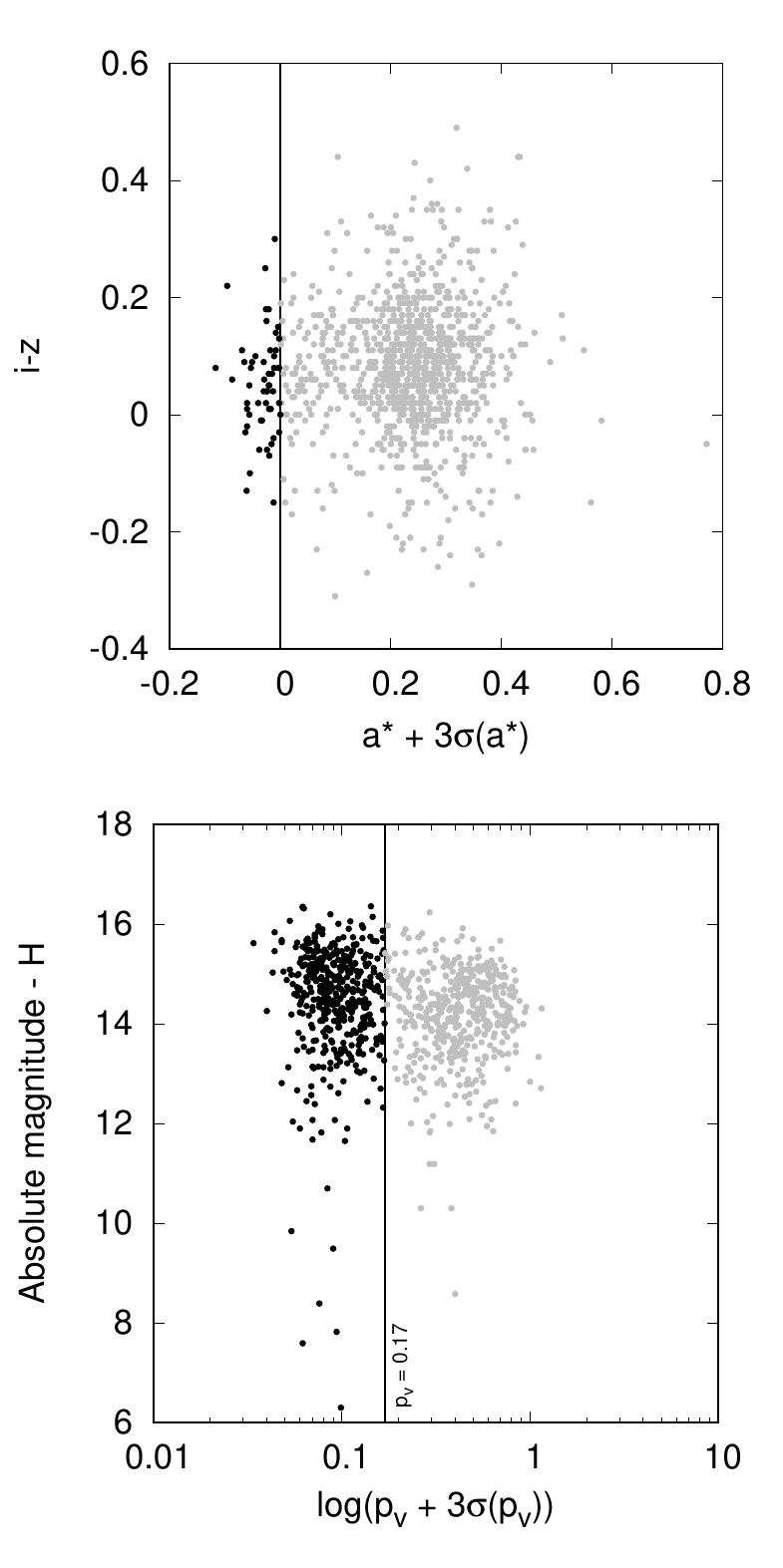}
 \caption{Identification of interlopers among the Klumpkea family members based on the SDSS colour (\textit{top panel}) and albedo (\textit{bottom panel}) data. Black and grey dots represent interlopers and family members, respectively.  The vertical lines show border between the real family members and interlopers. }
 \label{f:sdss}
\end{figure}

Being interested here in identifying potential interlopers among members of an asteroid family
we proceed in the following way. First, for each family, we calculate the mean value $a_{m}^{*}$ and its corresponding standard deviation $\sigma (a_{m}^*)$, using colours of potential family members.
The mean value is then used to define the spectral class of the family. Finally, all potential family members having their individual $a^*$ colour inconsistent with the one derived for the family are classified as interlopers. In order to keep the interloper identification as reliable as possible we introduce a safety 
margin between two types, based on the value of standard deviation of $a^{*}$ colour of an asteroid $\sigma (a^*)$. For instance, if an asteroid family is 
classified as $C$-type, only objects having ${a}^* < -3\sigma (a^*)$ are consider to be 
interlopers, and likewise, if a family is $S$-type then objects with ${a}^{*} > 3\sigma (a^*)$ are
interlopers. This methodology is illustrated in Fig~\ref{f:sdss} for the case of the Klumpkea family.\footnote{It should be noted that the Klumpkea family is listed as the Tirela family (FIN 612) in
\citet{nes2015} classification.}

\subsubsection{The geometric albedos - WISE, AKARI and IRAS}
\label{sss:albedo}

Another information that could be used to separate $C-$ from $S-$ taxonomic class, and is available for relatively large number of asteroids, is the geometric albedo. The vast majority of these data is 
obtained by three different infrared surveys. The catalogue of asteroid albedos provided by the Infrared Astronomical Satellite (IRAS) survey was presented by \citet{Tedesco2002}, and contains albedos and diameters for 2470 objects. The AKARI survey provides a catalogue with data 
for 5208 asteroids \citep[][]{Usui2011, Hasegawa2013}. Finally, most of currently known asteroid albedos are provided by the Wide-field Infrared Survey Explorer \citep[WISE;][]{Mainzer2011a,masiero2011}. Based on the data provided by this survey, 
\citet{masiero2011} published diameters and albedos for more than 100,000 Main-Belt asteroids.  

Aiming here to use these albedos to identify interlopers among potential family members, we
need to define strict criteria. In this respect, first a spectral type
of considered asteroid family should be determined, and then it should
be checked for each potential member if its own albedo is consistent with
the spectral type of the family. 

In order to determine the dominant
taxonomic class of a family, we simply calculate the average albedo $\overline{p^{*}_v}$ over all
members for which this information is provided by at least one of three available surveys\footnote{If the albedo is provided by more than one survey, we always give priority to the most recent data.}, and check if the obtained value is consistent with one of the spectral types. For this purpose we assume that an
albedo of $0.13$ is a border between two types, but, similarly as in the case of the SDSS data, we also introduced a safety margin. Therefore, an asteroid family is assigned $C$-type if $\overline{p^{*}_v} < 0.12$, while family is defined 
to be of $S$-type if $\overline{p^{*}_v} > 0.14$.

Most of known families fall in one of two broad spectral types defined above. However, as these two intervals do not overlap, in some cases this causes ambiguity in the determination of family spectral type. Therefore, for families with the average albedo in $[0.12, 0.14]$ interval, we used
predefined confidence intervals, suitable only for these families. Such an example is the Eos family (see Section~\ref{ss:other_fam}).

Once a spectral type is assigned to a family, all potential members of an
inconsistent spectral type are defined as interlopers. 
As a starting point we used results obtained by \citet{DeMeoCarry2013}. 
These authors calculated average 
albedo $\overline{p_v}$, and its corresponding standard deviation $\sigma(\overline{p_v})$, 
for different classes of asteroids, using data from three mentioned surveys.

Assuming that albedos of both spectral types ($C$ and $S$) do not spread more than 
$3\sigma$ from their average values, we calculated the corresponding confidence 
intervals. The average albedo and its corresponding standard deviation for 
$C$-type objects are $\overline{p_v}=0.06$ and $\sigma(\overline{p_v})=0.01$, respectively; 
thus, a confidence interval is $[0.03,0.09]$.
Similarly, $\overline{p_v}=0.23$ and $\sigma(\overline{p_v})=0.02$ for $S$-type,
resulting in a confidence interval of $[0.17,0.29]$. 

Finally, if a family is of $C$-type, all potential members with $p_v - 3 \sigma > 0.09$ are interlopers.
Accordingly, if a family belongs to the $S$-type, interlopers are all potential 
members having $p_v + 3 \sigma < 0.17$. For asteroid families with  predefined confidence interval, 
interlopers are those objects that have $p_v \pm 3 \sigma$ outside the corresponding interval.

The above defined criteria should be good enough even at small sizes, for which 
\citet{Mainzer2011b} found that the $S$-complex partially overlaps the low 
albedo $C$-complex.

\subsubsection{The spectroscopic data}

The different taxonomy classifications could also be used to exclude
interlopers from asteroid families. Unfortunately these data are available only
for a very limited number of asteroids. Still, this information should be
more reliable than e.g. SDSS colours. Therefore, we use them in our analysis.

Before we describe in details how individual interlopers are identified using spectroscopic data,
it is important to note that a family spectral class is defined based on the results obtained using
colour and albedo information. This is because the number of family members with available
spectra is often too small to reliably determine the dominant spectral class of the family.

Therefore, if, following the criteria described above, both, colour and albedo
data suggest that the family is either $C$ or $S$ class, we proceed with the exclusion of interlopers
using spectroscopic data. However, if these two sources of information are not in 
agreement\footnote{This may happen if a family belongs to $X$-type. It is because the SDSS
$a^{*}$ colour dose not separate $C$ from $X$ type, although these two types are characterized by 
quite different albedos.}, we completely omitted spectral data from our analysis. 

If applied, this analysis is based on \citet{tholen1984}, \citet{busbinzel2002} and \citet{demeo2009} classifications. 
As in the case of albedos, if more classifications provide
a spectral type of an asteroid, priority is given to the most recent data.

\citet{tholen1984} used spectrophotometric results from ECAS (Eight-Colour Asteroid
Survey) to classify asteroids in different taxonomy groups.   The main groups
are $C$, $S$ and $X$. Dark, carbonaceous asteroids are usually part of the $C$-group
and they could be divided in a few different classes by comparing slope and
maximum wavelength of its spectrum. In this respect, members of the $C$-group are $B$, $F$, $G$ and $C$
classes. Metallic asteroids are members of $X$ group that is divided into the $E$, $M$
and $P$ classes according to the asteroid geometric albedo. Siliceous-stony asteroids are members of the $S$-group. There
are some groups similar to $S$, but with slightly different spectrum slope
and maximum wavelength, such as $Q$, $R$, $V$ and $A$ classes. $D$ and $T$ classes can not be
easily classified in any of three broad groups defined by \citet{tholen1984}. If an
asteroid family is mostly composed of asteroids belonging to $C$ group, then 
an objects is an interloper if according to the Tholen's spectral classification it
belongs to one of the following classes: $S$, $E$, $M$, $P$, 
$Q$, $R$, $V$, $D$ and $T$. Analogously, if a family belongs to $S$ group, its potential 
member is an interloper if belongs to one of the following classes: $C$, $B$, $F$, $G$,
$D$, $T$ or $EMP$.

\citet{busbinzel2002} used data from Small Main-Belt Asteroid Spectroscopic
Survey (SMASS), which provided continuous spectrum, but covered smaller wavelength range then
the ECAS. This classification divided asteroids in 26 taxonomic classes, with most
of the classes being similar to those defined by \citet{tholen1984}, although some Bus \& Binzel classes are
separated into subclasses. For example, the $S$ group is separated in $S_a,
S_k, S_l, S_q, S_r$ and $S$ class, and hence, if a family is part of $C$ group, all asteroids that belong to the $S$ group are identified as interlopers. In this case we exclude as interlopers asteroids of $V, O, L, D, T, A, R, Q, K$ and $E$ classes. Similarly, if a family belongs to $S$ group, interlopers are those asteroids that belong to the $C$ group or $B, V, T$ and $D$ classes. 

A similar classification logic was applied in \citet{demeo2009} classification, but these authors used also the infrared part of the spectrum. Therefore, their classification is similar to that of Bus and Binzel, with some small modifications that eliminated subclasses $Ld$, $Sk$, and $Sl$, but added a new $Sv$ subclass within $S$ class. 

\subsection{STEP 3 and STEP 4}

\subsubsection{Excluding interlopers}

\begin{figure}
 \includegraphics [width=84mm]{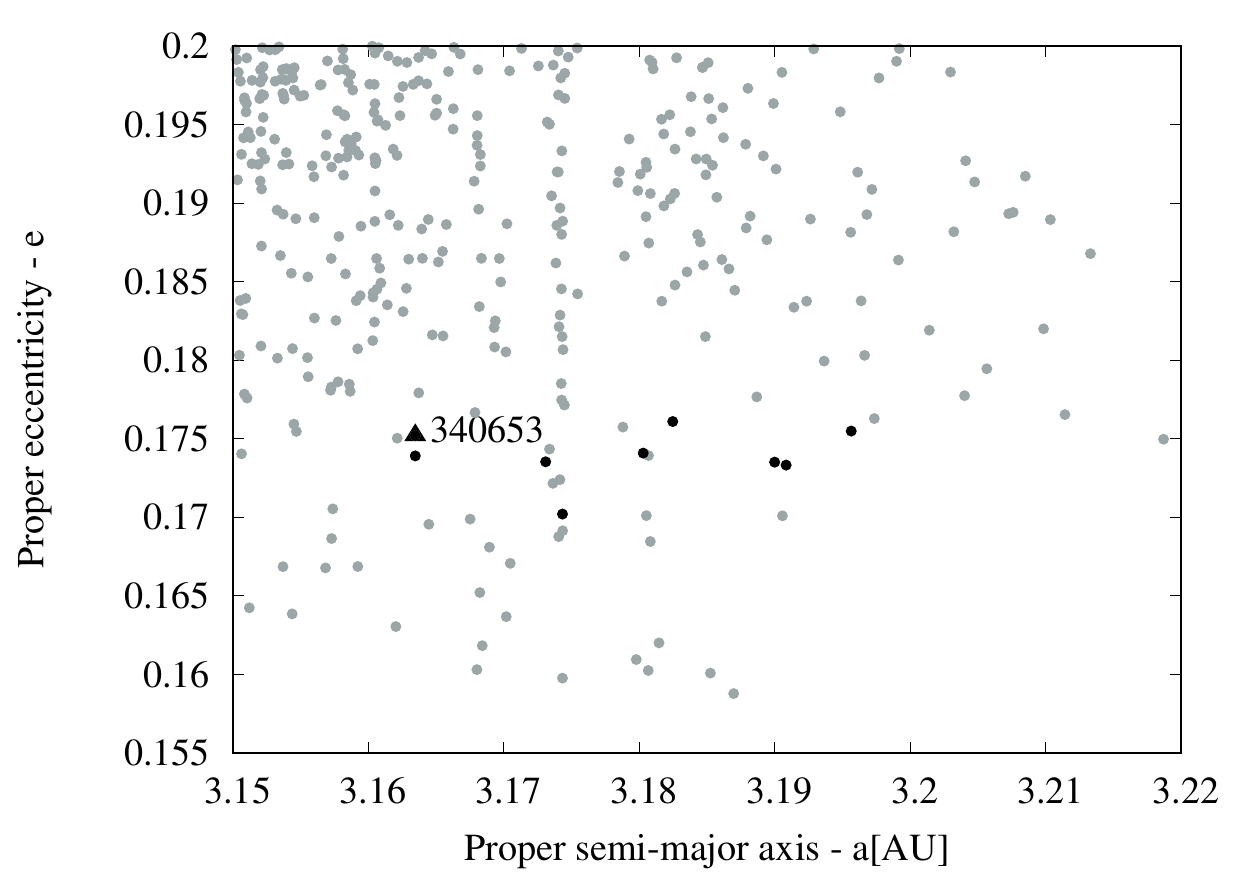}
 \caption{An example of the chaining effect in the Klumpkea family. Black triangle shows location of asteroid (340653) that is identified as an interloper, while black circles represent the eighth additional interlopers, linked to the family due to the chaining through the asteroid (340653). 
 Grey dots mark other family members. }
\label{f:klumpkea_chaining}
\end{figure}
 
In the third step, asteroids identified as interlopers due to their reflectance characteristics are excluded from the initial catalogue of proper elements.
This produces a modified catalogue, that is free of known interlopers in the corresponding family. Therefore, the interlopers are excluded from the catalogue of proper elements, rather than from the list of family members, as it is usually done.

In the fourth step the HCM is applied again to the modified catalogue of proper elements, produced in the third step. This is the main difference from the methods that have already been used for determining interlopers in asteroid families. The point of this step is to reduce the chaining effect in the HCM. By removing interlopers from the catalogue, we have also removed objects linked to the family through some of these interlopers. An example of the efficiency of this methodology in the Klumpkea family is shown in Fig~\ref{f:klumpkea_chaining}. Asteroid (340653) is identified as an interloper based on the albedo data. Once it is removed from the input catalogue, the eighth additional objects are removed.

\subsubsection{The final membership}

Finally, in the fourth step, the family members are re-identified using the modified catalogue of proper elements, produced in the third step
by removing interlopers identified based on colour, albedo and spectral data. For a purpose of comparison, we run the HCM on the modified catalogue 
using again the cut-off distance adopted in the first step. This allows to estimate the fraction of removed interlopers within each family.

On the other hand, as discussed above, the cut-off used in the first step is too high to obtain the nominal membership of
an asteroid family. A large cut-off value may easily associate many background asteroids to a family. This may even create bridge between different families resulting in a wrong membership. Therefore, the final membership of each family should be obtained using the cut-off that describes a family the best on the modified catalogue of proper elements. Hence, the determination of a nominal cut-off is a very important step in this analysis, and should be carefully performed. 

The method used here to select the nominal cut-off value is as follows.
First, it is checked how the number of family members changes as a function of cut-off values (Fig.~\ref{f:klumpkea_hcm}). Then, in order to estimate a nominal cut-off we search for a \textit{plateau}, i.e. an interval of cut-off values where a number of family members is almost constant. As a general rule, if such a plateau is well defined, the nominal cut-off is adopted to be around the centre of the plateau. Still, in some cases the plateau practically does not exist, and a detailed analysis of  corresponding family should be performed in order to determine its nominal cut-off value. 
The latter cases are discussed below, in Section~\ref{ss:other_fam}.

\section[]{Results}

\begin{table*}
 \centering
 \begin{minipage}{180mm}
  \caption{Summary of results on excluding interlopers from selected families. The columns are: (1) Family name (Family), (2) Adopted cut-off value ($d_{cut}$), (3) Number of asteroids as family members identified using cut-off values adopted in step 1 (\# STEP1), (4) Number of interlopers identified using SDSS colours (\# SDSS), (5) Number of interlopers identified using albedos (\# ALBEDO), (6) Number of interlopers identified using available taxonomic classifications (\# TAX.), (7) Total number of interlopers based on the reflectance spectra (\# STEP2), (8) Number of asteroids excluded due to the chaining effect (Chaining), (9) Number of asteroids identified as family members using adopted $d_{cut}$ and modified input catalogue (\# STEP 4), (10) A fraction of excluded asteroids (\%) }
    \label{t:results01}
   \centering
  \begin{tabular}{@{}c|ccccccccc@{}}
  \hline
  Family & $d_{cut}$ & \# STEP1 & \# SDSS & \# ALBEDO & \# TAX. & \# STEP2 & Chaining & \# STEP 4 & \% \\
 \hline
 (5) Astraea 	& 55 & 7482		& 92 	& 295 	& 2  & 361 	& 234 	& 6887 	& 7.9	\\
 (10) Hygiea 	& 70 & 5904		& 24 	& 15 	& 3  & 38 	& 86 	& 5780 	& 2.1	\\
 (15) Eunomia 	& 60 & 11889	& 411 	& 1421 	& 12 & 1595	& 316 	& 9978 	& 16.1	\\
 (20) Massalia 	& 30 & 4663 	& 7 	& 8 	& 0  & 13 	& 2 	& 4648 	& 0.3 	\\
 (24) Themis 	& 80 & 5499 	& 20 	& 39 	& 0 & 59 	& 31 	& 5409 	& 1.6	\\
 (93) Minerva 	& 75 & 7015 	& 352 	& 845 	& 39  & 1057& - 	& - 	& -		\\ 
 (135) Hertha 	& 45 & 22849 	& 465 	& 1138 	& 10  & 1358& 1363 	& 20128 & 11.9	\\
 (145) Adeona 	& 45 & 1994 	& 45 	& 21 	& 0  & 62 	& 78 	& 1854 	& 7.0\\
 (158) Koronis 	& 65 & 7743 	& 28 	& 81 	& 1  & 101 	& 38 	& 7604 	& 1.8 \\
 (170) Maria 	& 60 & 2939 	& 20 	& 28 	& 0  & 45 	& 44 	& 2850 	& 3.0\\ 
 (221) Eos 		& 70 & 24155 	& 555 	& 1706 	& 39  & 2089& 757 	& 21309 & 11.8\\
 (490) Veritas 	& 30 & 1295 	& 6 	& 7 	& 0  & 13 	& 0 	& 1282 	& 1.0\\
 (668) Dora 	& 60 & 1401 	& 9 	& 3 	& 1  & 13 	& 0 	& 1388 	& 0.9\\
 (847) Agnia 	& 45 & 3054 	& 14 	& 52 	& 0  & 61 	& 84 	& 2909 	& 4.7\\
 (1040) Klumpkea	& 80 & 2794	& 56 	& 435 	& 3  & 452 	& 227 	& 2115 	& 24.5\\
 (1726) Hoffmeister	& 40 & 1763	& 6 	& 3 	& 1  & 9 	& 0 	& 1754 	& 0.5\\
 (2076) Levin 		& 45 & 2500	& 52 	& 30 	& 2  & 83 	& 71 	& 2346 	& 6.2\\
 \hline
\end{tabular}
\end{minipage}
\end{table*}

In this section we demonstrate how the above described method works, and present the most important results. 
For this purpose we selected 17 large families from the classification of \citet{milani2014}, 
each of them with more than 1,000 potential members. These families are statistically 
reliable, but at the same time are also expected to have a large number of interlopers.

Before discussing these results, let us to recall that they are obtained using unrealistically
high velocity distance cut-off values ($d_{cut}$). The purpose here is only to show
how the method works, and in particular to highlight the importance of our step \#4.
Realistic results, obtained using the most appropriate values of $d_{cut}$, are given below
in Section~\ref{ss:other_fam}.

The results for the 17 analysed families are summarized in Tables~\ref{t:results01} and~\ref{t:results_average}. The total fraction of interlopers found among potential family members 
vary from below 1\% for families well separated from the background population (e.g. the Hoffmeister family), to more than 20\% for cases where an additional family may be present, as for example
in the case of the Klumpkea family. Despite these differences, the results are generally in agreement with the expected fraction of interlopers estimated by \citet{migliorini1995}.

\begin{table}
 \centering
 \begin{minipage}{90mm}
   \centering
  \caption{Average colour and albedo characteristics of analysed families. The columns are: (1)  Family name (Family), 
  (2) average SDSS $a^*$ colour ($a_{m}^{*}$), (3) standard deviation of average $a^*$ colour ($\sigma_{m} (a^*)$),
  (4) average albedo ($p^{*}_v$), (5) standard deviation of average albedo ($\sigma (p^{*}_v)$), 
  and (6) Determined family type (Type) }
    \label{t:results_average}
  \begin{tabular}{@{}c|ccccc@{}}  
  \hline
  Family & $a_{m}^{*}$ & $\sigma_{m} (a^*)$ & $p^{*}_v$ & $\sigma (p^{*}_v)$ & Type  \\
 \hline
 (5) Astraea 		& 0.09 	& 	0.03 	& 0.19  & 0.06 	& S\\
 (10) Hygiea 		& -0.10 & 	0.03 	& 0.07  & 0.02	& C\\
 (15) Eunomia 		& 0.08  & 	0.03 	& 0.20  & 0.05	& S\\
 (20) Massalia 		& 0.07 	& 	0.04   	& 0.24  & 0.09 	& S\\
 (24) Themis 		& -0.11 & 	0.03   	& 0.07  & 0.02	& C\\
 (135) Hertha 		& 0.08 	&	0.03	& 0.17 	& 0.06  & S \\ 
 (145) Adeona 		& -0.09 & 	0.04 	& 0.07 	& 0.02 	& C\\
 (158) Koronis 		& 0.09 	& 	0.03  	& 0.23 	& 0.07	& S\\ 
 (170) Maria 		& 0.11 	& 	0.03  	& 0.25 	& 0.08 	& S\\
 (221) Eos 		& 0.03 	& 	0.03  	& 0.13 	& 0.04 	& S\\
 (490) Veritas 		& -0.07 & 	0.03 	& 0.07 	& 0.02 	& C\\
 (668) Dora 		& -0.11 & 	0.03 	& 0.06 	& 0.02	& C\\
 (847) Agnia 		& 0.07 	& 	0.03  	& 0.19 	& 0.07	& S\\
 (1040) Klumpkea 	& 0.11 	& 	0.03  	& 0.14	& 0.04 	& S\\
 (1726) Hoffmeister	& -0.09 & 	0.03 	& 0.05 	& 0.02 	& C\\
 (2076) Levin		& 0.04 & 	0.03 	& 0.21 	& 0.07 	& S\\
\hline
\end{tabular}
\end{minipage}
\end{table}

To better appreciate the importance of step \#4, which is the main improvement of our approach with
respect to previous works, one should compare the numbers shown in columns 7 and 8 of Table~\ref{t:results01}. 
The 7th column gives the total number of interlopers identified
using available physical and spectral data (\# STEP2), while the 8th column provides the number of
asteroids linked to the family through some of the identified interlopers (Chaining).

The numbers in these two columns are often comparable, and in some cases the number of
interlopers removed due to chaining effect is even larger then the number of interlopers
identified based on the physical and spectral data. The only exceptions are families with
a very small number of interlopers (e.g. Veritas and Hoffmeister), and cases where two or
more families overlap (e.g. Eunomia family). Nevertheless, usage of step \#4 seems to be
important and fully justified.

In order to further demonstrate how the method works we selected two families as the case studies,
namely the (1040)~Klumpkea and (15)~Eunomia families. The Klumpkea family was selected due to
its location within the main-belt. It is situated in the outer part of the main-belt, 
where most of the asteroids belong to $C$ spectral type, while Klumpkea belongs to $S$-type. 
Therefore, we expect to find a relatively large fraction of the interlopers among members of the Klumpkea family. 
The Eunomia family is a large group in the middle main-belt, and it is an 
interesting example because there may be another family buried inside the Eunomia family.

\subsection{The case study \#1: Klumpkea family}

\begin{figure}
 \includegraphics [width=84mm]{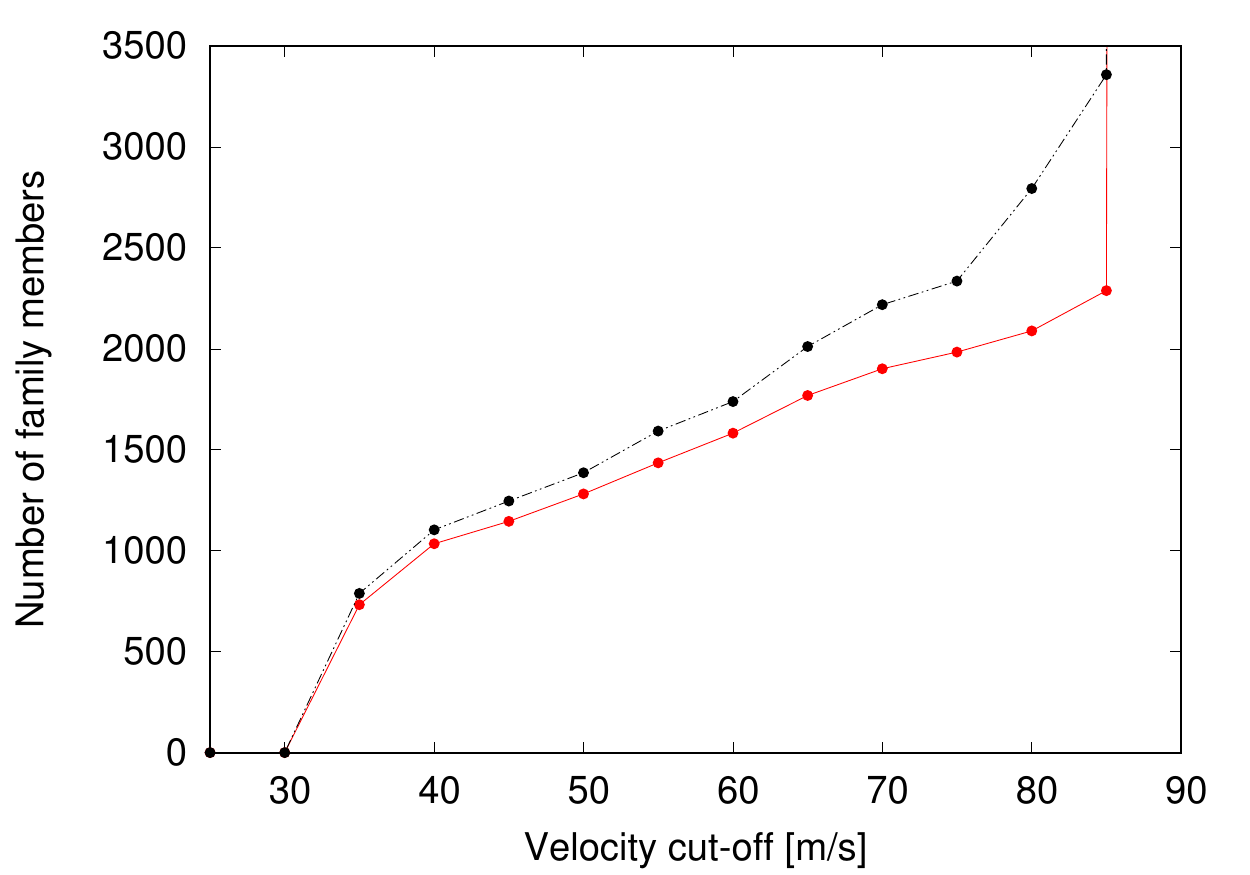}
 \caption{Number of asteroids linked to the Klumpkea family. The dashed and solid line represents
 data obtained using initial and modified catalogue respectively. }
\label{f:klumpkea_hcm}
\end{figure}

Analysing the outputs of the HCM applied to the initial catalogue for different cut-off values, 
it could be noted that for cut-off below $55~ms^{-1}$, Klumpkea family members are situated 
between the 11/5 and 21/10 mean motion resonances with Jupiter. Starting from $55~ms^{-1}$ family members are crossing the 21/10 resonance, while at $80~ms^{-1}$ family members cross also the 11/5 resonance. At the cut-off velocity of $90~ms^{-1}$, the family merges with the background population of asteroids. Therefore, the cut-off value of $80~ms^{-1}$ is adopted for the step \#1, yielding
the initial family membership that contains 2,794 asteroids (see Fig.~\ref{f:klumpkea_hcm}). 
These potential members of Klumpkea family have an average albedo $\overline{p_v}= 0.142 \pm 0.044$, and an average SDSS colour $\overline{a^*} = 0.112 \pm 0.035$. Hence, according to the criteria explained in Section~\ref{ss:interlopers_identification} the family belongs to the $S$-type.

\begin{figure*}
 \includegraphics [width=170mm]{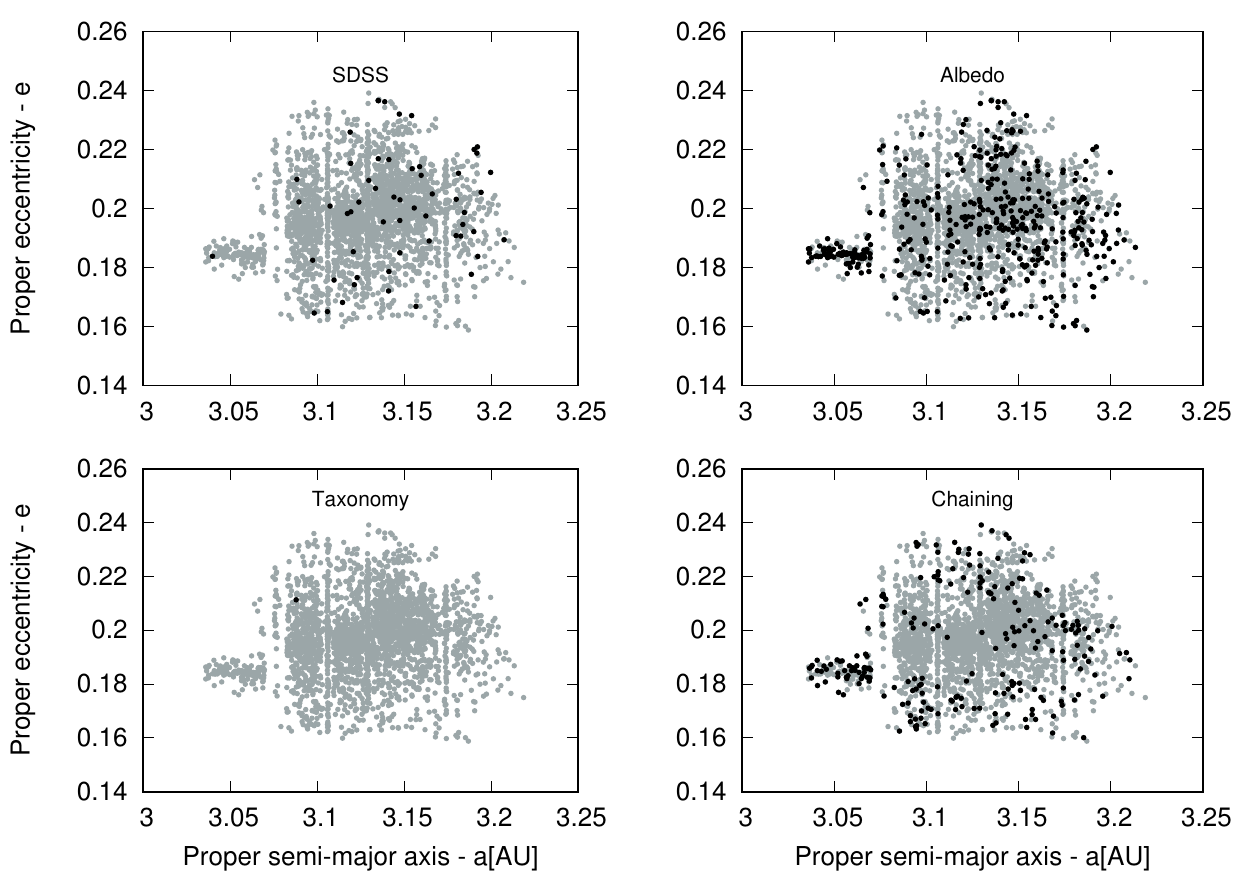}
\caption{Distribution of interlopers within the Klumpkea family in the $a - e$ plane. Each panel shows
interlopers identified using different information, as written in the panels. Black and gray circles mark interlopers and all asteroids linked to the family, respectively.}
\label{f:klumpkea_interlopers}
\end{figure*}

In the second step, among the initial family members 452 interlopers are detected (see Table~\ref{t:results01}). Fig.~\ref{f:klumpkea_interlopers} shows the distribution of these interlopers in 
different planes. It should be noted that most of the potential members located at $a < 3.075$~au are
likely interlopers.

After removing the interlopers from the initial catalogue, and applying the HCM to the modified catalogue\footnote{Using the same $d_{cut}$ as in the first step}, 2,115 asteroids were identified
as family members. Therefore, there are 679 members less than in the initial step, and among 
these 227 asteroids are removed in step \#4 due to the chaining effect.  
 
\begin{figure*}
\includegraphics [width=170mm]{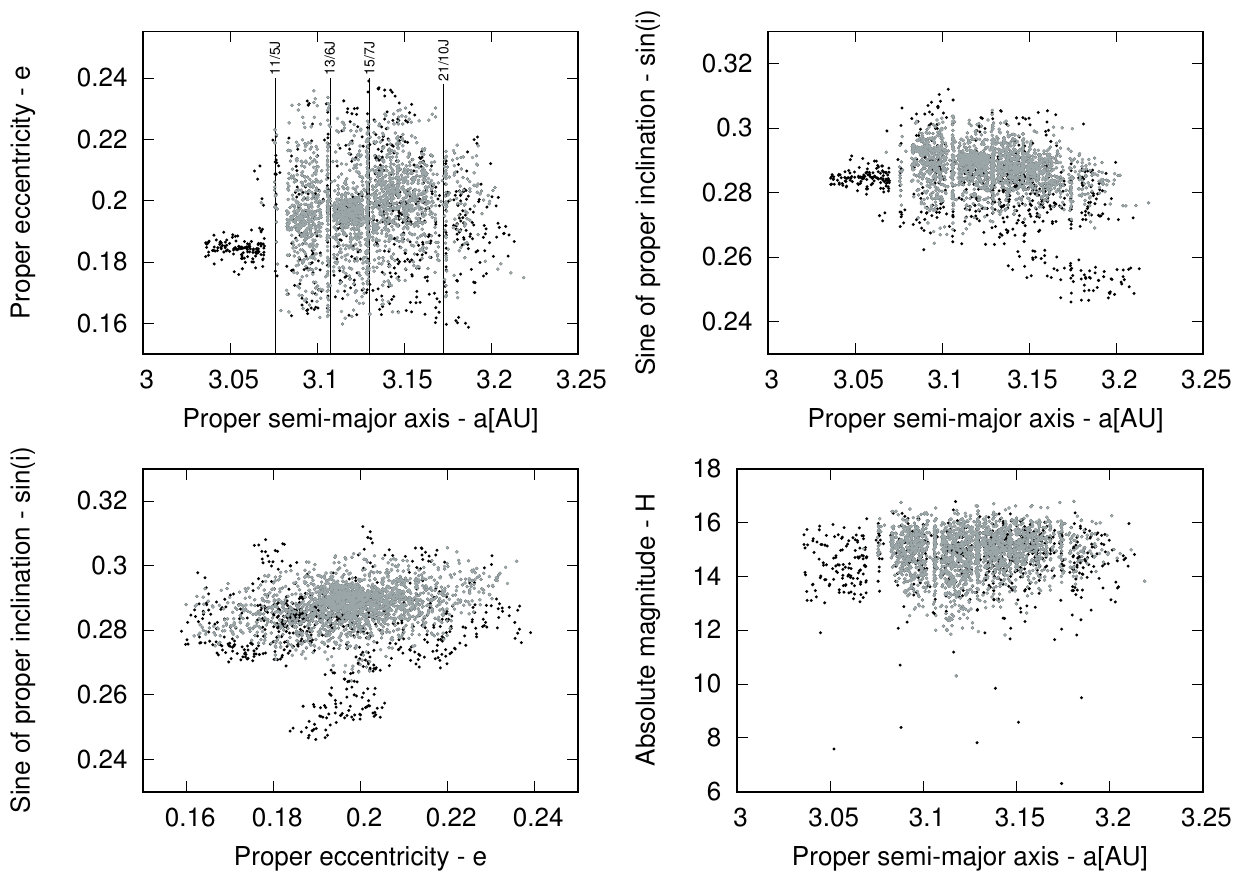}
 \caption{The final membership (grey) and all identified interlopers (black) of the Klumpkea family,
 shown in the space of proper elements and absolute magnitude. The vertical lines in the top-left panel denote locations of the most important mean-motion resonances that cross the family.}
 \label{f:klumpkea_final}
\end{figure*}

In order to determine the nominal cut-off for the Klumpkea family, we study again the HCM results,
but this time obtained on the modified catalogue. The situation seems to be different with respect
to the results obtained before removal of interlopers. The most important new feature is that family members do not cross the 11/5 resonance, for cut-off values below $85~ms^{-1}$ when the family  
merges with the local background population (Fig.~\ref{f:klumpkea_final}). These results are consistent with two recent classification performed by \citet{milani2014} and \citet{nes2015}, where the authors found that
at the inner side of the 11/5 resonance there is another family, namely that of (96) Aegle. 

Moreover, a low inclination part of the initial family is completely removed (see upper-right
panel of Fig.~\ref{f:klumpkea_final}). Finally, it is interesting to note that in the $a-H$ plane
(lower-right panel of Fig.~\ref{f:klumpkea_final}) most of the objects located below the V-shape are identified as interlopers, and the removal of these asteroids improve the visibility of the family
V-shape.\footnote{The so-called V-shape is a characteristic shape of the real collisional families when projected on the semi-major axis vs. absolute magnitude (or inverse diameter) plane. It is mainly the result of the action of Yarkovsky thermal force which disperse more quickly smaller objects \citep{vok2015}. The shape structure is mostly used to estimate the age of a family \citep[][]{vok2006,milani2014,spoto2015}, but could also be used to search or confirm existence of asteroid families \citep{walsh2013,bolin2017}.}

For the nominal cut-off value of the Klumpkea family we adopt $d_{nom} = 60~ms^{-1}$, 
that corresponds to the removal of 9.8\% of asteroids initially linked to the family. 
About 18\% of interlopers are removed in the fourth step of our procedure, that is introduced
to reduce the chaining effect.

\subsection{The case study \#2: Eunomia family}

\begin{figure}
 \includegraphics [width=84mm]{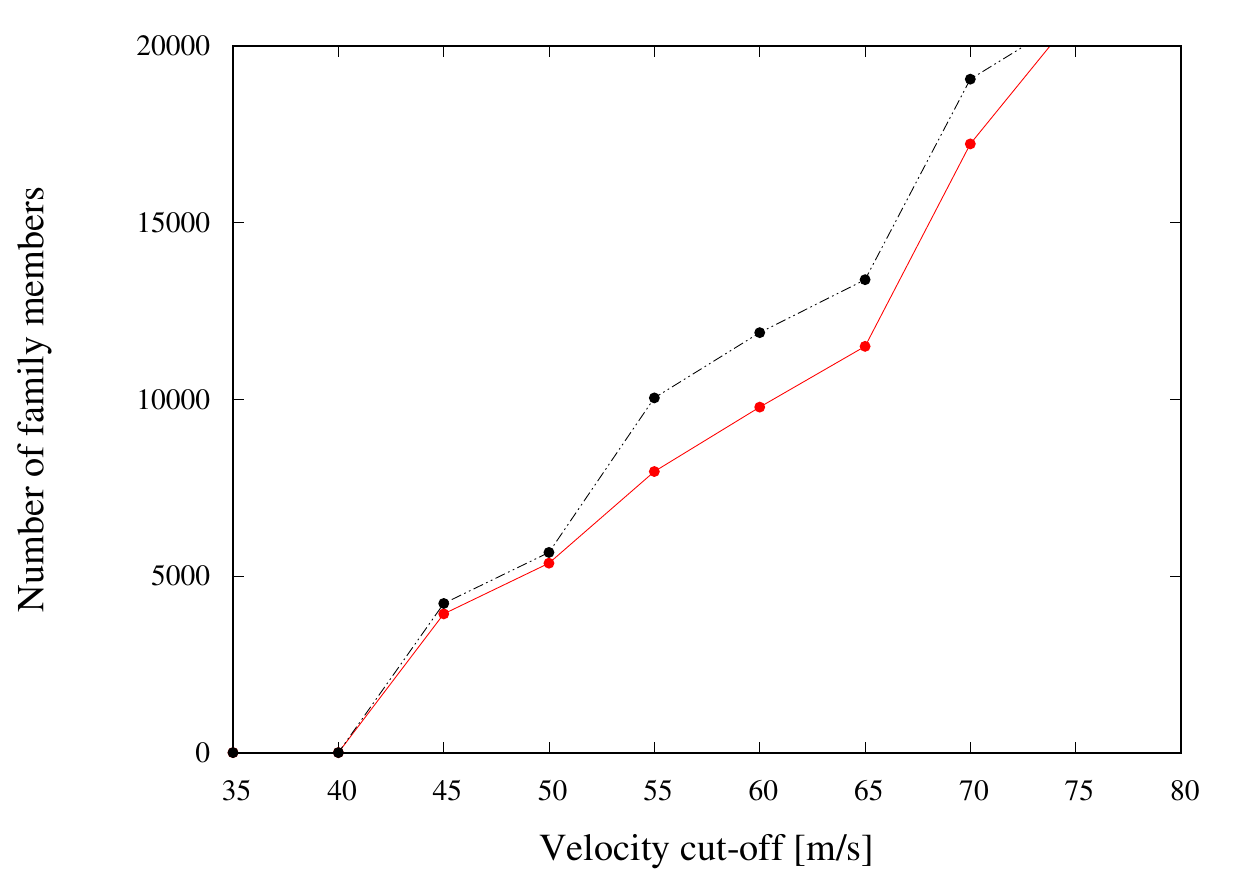}
 \caption{Number of asteroids linked to the Eunomia family. The dashed and solid line represents
 membership obtained using initial and modified catalogue respectively.}
 \label{f:15_hcm}
\end{figure}

To get the initial list of family members we start from the cut-off value of $5~ms^{-1}$ and increase it until the family merges with background asteroids (Fig.~\ref{f:15_hcm}). Based on the defined criteria we adopt a cut-off value of $60~ms^{-1}$, resulting in 11,889 potential members. 

The above defined Eunomia family has an average albedo $\overline{p_v}= 0.199 \pm 0.056$, and average SDSS colour $\overline{a^*} = 0.084 \pm 0.030$, suggesting that this family belongs to the $S$-type.

\begin{figure*}
 \includegraphics[width=170mm]{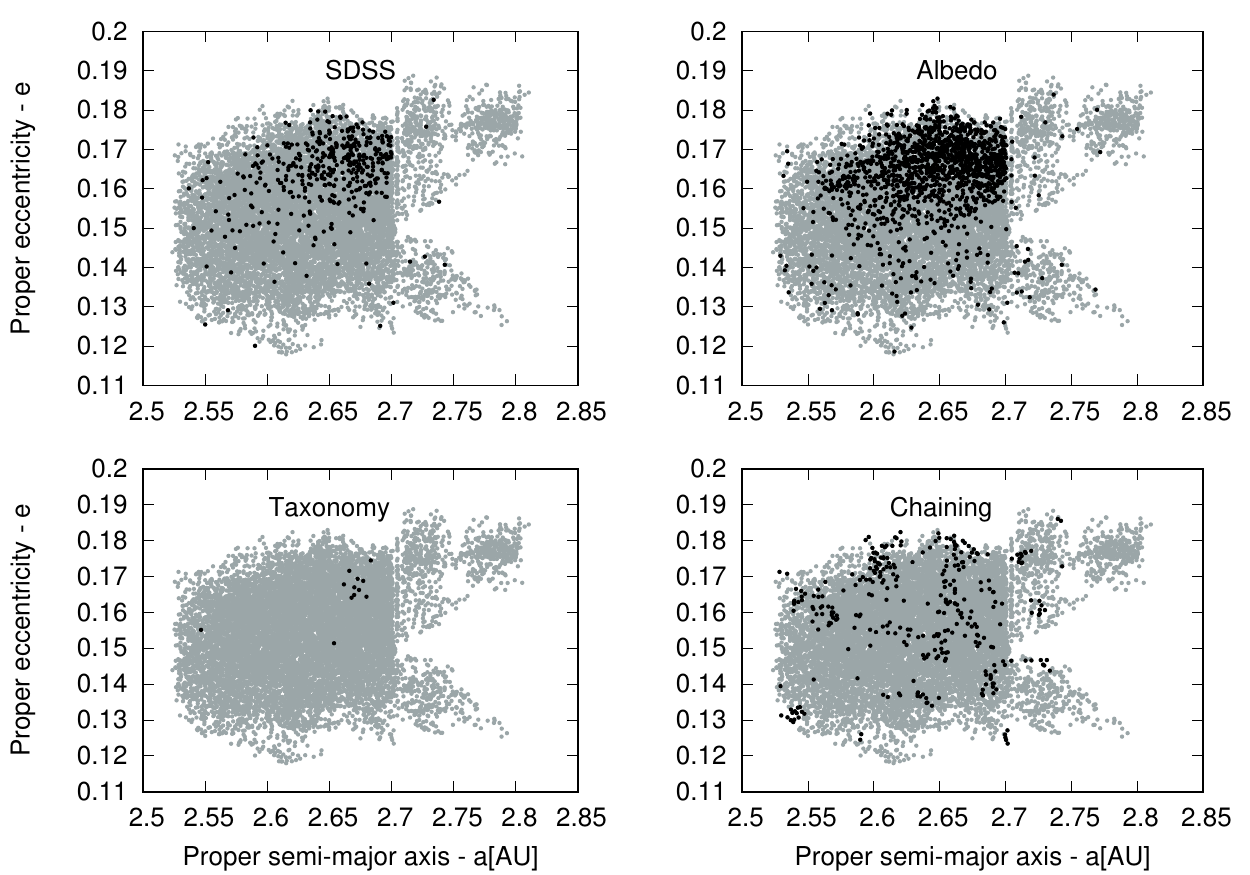}
\caption{Distribution of interlopers within Eunomia family in the $a - e$ plane. The grey points represent family members, while the black points mark interlopers. }
\label{f:15_interlopers}
\end{figure*}

\begin{figure*}
\includegraphics [width=170mm]{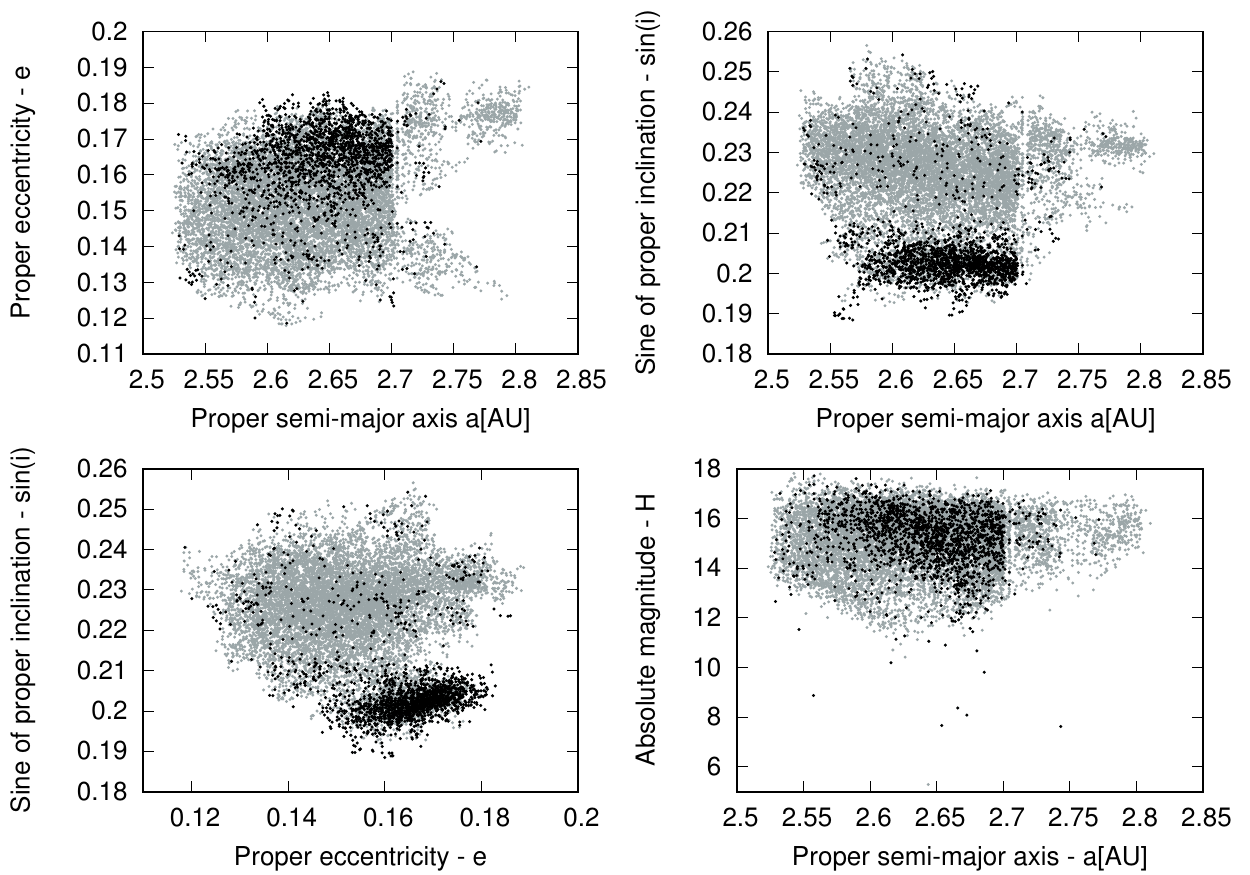}
 \caption{The final membership (grey) and all identified interlopers (black) of the Eunomia family. }
 \label{f:15_final}
\end{figure*}

In the preliminary list of members we found 1,595 interlopers (see Table~\ref{t:results01} for more details), whose distributions are shown in Fig.~\ref{f:15_interlopers}. Applying the HCM with the same $d_{cut}$ as in the first step to the modified catalogue, 9,978 asteroids are identified as the family members (Figs.~\ref{f:15_interlopers} and \ref{f:15_final}). Therefore, there are 1,911 (about 16\%) asteroids linked to the family less than in the initial step, with 316 of them rejected in step \#4. 
 
A cut-off value of $55~ms^{-1}$ is adopted as a nominal to define the family, resulting
in removal of almost 20\% of asteroids linked to the family as interlopers. \citet{migliorini1995} analyzed the Eunomia family and found that a large fraction of interlopers $\sim 20$\% should indeed be expected in this family, in agreement with the results obtained in our work. A possible explanation for such a large fraction of interlopers may be the presence of another asteroid family. Additional argument in favour of this hypothesis is concentration of interlopers in specific part of the Eunomia family (see Fig.~\ref{f:15_interlopers}). Examining the available data we found that $C$-type Adeona should be this overlapping family, as its location corresponds to the groupings of interlopers visible within the borders of the Eunomia family.  

Examining the V-shape structure of the Eunomia family shown in Fig.~\ref{f:15_final}, we noted that it is a very well defined, and practically all objects positioned outside the V-shape are identified as interlopers. Moreover, even the interlopers form a visible V-shape structure,
although cut at outer side in terms of the semi-major axis. This is an additional evidence that most of the interlopers belong to another collisional family, i.e. to the Adeona family.

The results of this case study illustrate the efficiency of the method developed here,
when dealing with two overlapping families of different spectral types.

\subsection{Special cases}
\label{ss:special_cases}

The special cases presented in this subsection refer to groups composed of two or more overlapping collisional families. Aiming to keep our approach fully automatic and relatively 
simple, obviously we do not expect to completely solve these complex cases. Moreover, as we are distinguishing here only between the $C$- and $S$- taxonomic complex, we could not resolve 
more than two overlapping families. Therefore, the main purpose of this part is to demonstrate 
how our method works in these cases, rather than to provide a full explanation for them.

\subsubsection{Minerva clan}

The asteroids from the regions around (1)~Ceres are strongly perturbed by the nearby secular resonances with Ceres \citep{nov2015,tn2016}. The existence of these resonances complicates any attempt to identify asteroid families.

Asteroid Minerva was first considered as a part of Ceres dynamical family \citep{Zappala1995}. The later works also recognized this group, but its largest associated object has changed. It was proposed that asteroid Ceres is not a member of this group\footnote{For a discussion on the missing Ceres family see \citet{milani2014}, \citet{rivkin2014} and \citet{carruba2016}.} making (93) Minerva the largest member, and consequently the group is named Minerva family. However, later on it was realized that this family mainly consists of $S$-type asteroids, indicating that $C$-type Minerva asteroid is actually an interloper. Thus, the family name has changed once again to Gefion family. 

As \citet{milani2014} identified the so-called dynamical families,\footnote{Dynamical families are groupings of asteroids identified using purely dynamical characteristics, i.e. proper orbital elements. Therefore, they may not necessarily represent collisional families.} their classification includes the Minerva family. Therefore, we applied our analysis to this group, but note that asteroid (93) Minerva may not be a parent body of this group.

In the initial step, applied using  cut-off of $75\ ms^{-1}$ and asteroid (93)~Minerva as central object, we identified a group of 7,015 members. As this group includes more than one family (see discussion below), we refer to it as Minerva clan \citep{cellinozappala1993}. In the second step of our procedure applied to the Minerva clan we calculated the average colour: $\overline{a_*} = 0.044\pm0.032$ and albedo: $\overline{p_v} = 0.137 \pm 0.04$. Both these values are somewhere in between the typical values for $S$ and $C$ type. While according to our criteria colour data suggest the group should be marginally classified as $S$-type, the albedo value falls exactly between the values used to discriminate between $C$ and $S$ type. This is an indication that the Minerva clan includes asteroids of different composition and possibly more than one single family. This is further supported by the fact that the Minerva clan, as identified here, includes asteroids (1)~Ceres, (668)~Dora and (1272)~Gefion.

Due to existence of asteroids of different taxonomic types, we continue in two directions; that is, we first perform analysis assuming that the Minerva clan is $S$-type, but then repeated this step assuming it is $C$-type. In the first case, we found 1,497 interlopers, objects not compatible with $S$-type taxonomy. Among them there are asteroids (1), (93) and (668). After removing the interlopers from the initial catalogue, we obtained the modified catalogue of proper elements. Because asteroid (93)~Minerva is not present in the new catalogue, we applied the HCM analysis using (1272)~Gefion as a central body. For nominal cut-off value in this case we adopted $d_{nom}=50\ ms^{-1}$, and obtained well-defined Gefion family with 2,306 members. 

For the second case we assume Minerva clan is $C$-type and, in step \#2, identified 985 asteroids as interlopers, including (1272)~Gefion. Again, we exclude them from the initial catalogue, and apply the HCM using nominal cut-off of $65~ms^{-1}$ and asteroid Minerva as a central body. To this group of 3,184 asteroids we refer here as Minerva group. Our results show that difference between the initial and the final Minerva group is 1,356 asteroids. This large fraction of interlopers could be explained by the presence of Gefion family members, which we identify as separate family composed of $S$-type asteroids. 
 
\begin{figure}
 \includegraphics [width=84mm]{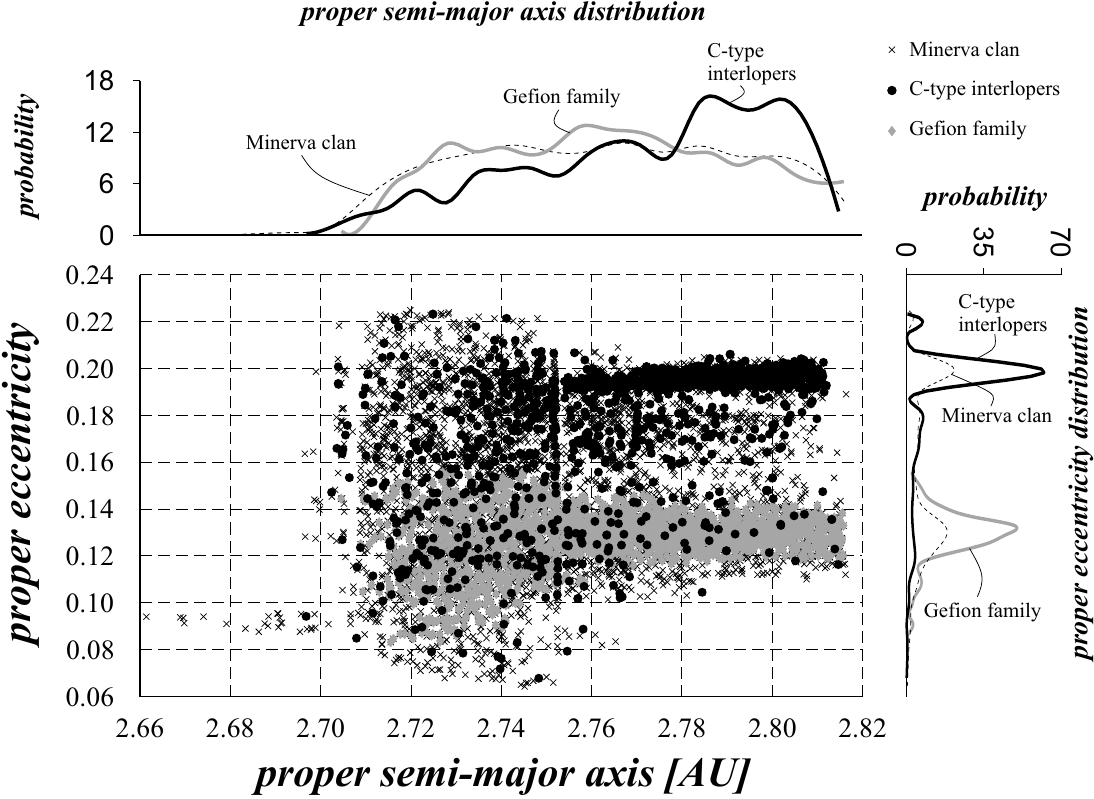}
\caption{The asteroids linked to the Minerva clan, along with its $C$-type interlopers and the Gefion family. The large concentration of dark interlopers visible in the upper-right part of the panel  corresponds to the Dora family. Note that apart from this grouping, the distribution of $C$-type interlopers is almost uniform, suggesting that dark Minerva family actually dose not exist. See text for additional explanation. In the top and right small panel probability density functions of the proper elements are given. These functions are obtained by interpolation of the normalized histograms.}
  \label{f:minervaclan}
\end{figure}

As shown in Fig.~\ref{f:minervaclan}, the Minerva group and Gefion family, as defined here, partly overlap in the proper elements space. In order to better understand the real nature of the Minerva group, we analysed the distribution of the family members in orbital space. We note that the largest concentration in the Minerva group coincides with the location of the Gefion family. On the other hand the distribution of the dark asteroids in this region is roughly uniform (Fig.~\ref{f:minervaclan}),
except at the location of the Dora family that was included in the membership of the Minerva group. Based on this evidence, we concluded that asteroids belonging to the Minerva group that coincide with the location of the Gefion family, are actually the members of the latter family for which albedo and/or colour data are not available. 

The above mentioned facts imply that the Minerva dynamically family is not a collisional family, while nearby there is one bright (Gefion) and one dark (Dora) family. Still, caution is needed before any
definite conclusion is derived about the families in this region. The secular resonances with Ceres complicate any attempt to identify asteroid families, and instead of the HCM, other techniques should be used to search for potential families, as explained in \citet{carruba2016}.
 
\subsubsection{Hertha/Nysa-Polana clan}

The Hertha family is one of the largest families that is considered in this paper. Other paper usually referred to it as Nysa-Polana complex of asteroids. This complex was first proposed as a single family by  \cite{Brouwer1951}, but later data showed that there are two or more overlapping collisional families \citep{cellino2001,walsh2013,dykhuis2015}. As this group is known to contain multiple families we will refer to it as Hertha clan. 

\begin{figure}
	\includegraphics [width=84mm]{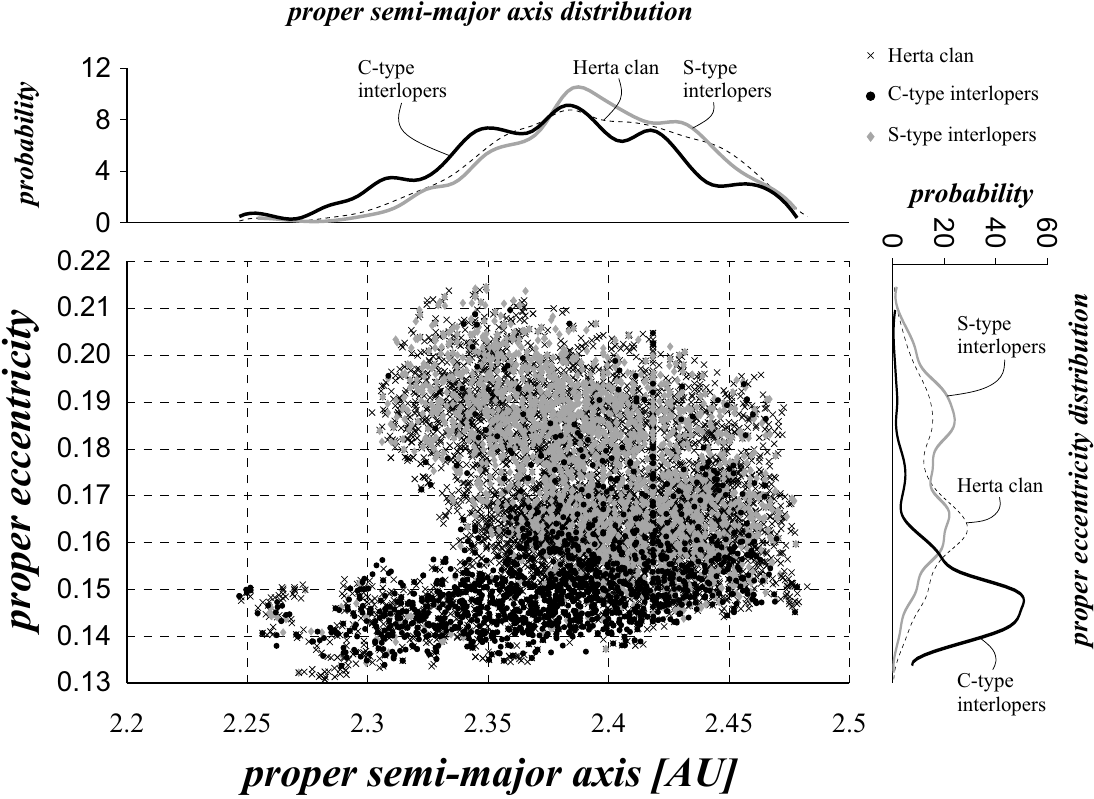}
	\caption{The $C$ and $S$-type interlopers within the Hertha clan. The distribution of these interlopers is clearly not random, but shows concentration around different values of proper eccentricity, indicating a presence of more than a single collisional family. In the top and right small panel probability density functions of the proper elements are given. A small offset in terms of proper semi-major , between the main concentrations of $C$- and $S$-type interlopers, is notable (top panel), but these two groups are obviously much better separated in terms of proper eccentricity (see right panel).}
	\label{img:hertha}
\end{figure}

Following our approach, in the first step by using the cut-off value of $45ms^{-1}$ 22,851 asteroids are identified as members of the Hertha clan (Figure~\ref{img:hertha}). According to the described criteria, the Hertha clan should be classified as $S$-type. Interestingly, the average colour and albedo do not show any indication that there may exist a family of different spectral type (see Table~\ref{t:results_average}). This may be because among the objects belonging to the Hertha clan, the $S$-type asteroids with available spectral data are more numerous than the $C$-type ones.

Being similar to the Minerva clan, we employed the same approach here in order to define families within the Hertha clan, i.e. we did not use only the spectral type of the clan derived in 
the second step, but run the procedure twice, once assuming the family is $S$-, and once
assuming it is $C$-type. 

In the first run we identified 1,603 $C$-type asteroids as interlopers, including asteroid (142)~Polana. After excluding $C$-type interlopers from the initial catalogue of proper elements,
the HCM analysis is performed using modified catalogue. For a nominal cut-off of $35ms^{-1}$ 
9,815 asteroids are associated to the family, with asteroid (135)~Hertha as its
the lowest numbered member. Therefore, after excluding
$C$-type asteroids, the Hertha family is obtained.\footnote{The second largest asteroid in the Hertha family identified in this work is (878) Mildred. \cite{cellino2001} excluded asteroid (135)~Hertha from its namesake family due to difference in spectrum of (135)~Hertha and the rest of family members. Therefore, these authors used asteroid (878) as the parent body of the family which they called Mildred family. As our conservative criteria did not exclude asteroid (135)~Hertha from the family membership, we still called it the Hertha family. } Asteroid (44) Nysa is included in
the Hertha family, but at a larger cut-off value of $45ms^{-1}$. 
Interestingly, \cite{dykhuis2015} proposed that the Hertha family consists of two sub-families, namely Hertha~1 and Hertha~2, however our methodology is not able to provide any evidence in this respect.

In the second run, 2,537 $S$-type asteroids are identified as interlopers, and among these
there are (20)~Massalia, (44)~Nysa and (135)~Hertha asteroids.
Therefore, in order to obtain $C$-type family within the Hertha clan, these $S$-type asteroids are excluded as interlopers. After doing this, asteroids (135)~Hertha and (44)~Nysa no longer belong to the Hertha clan. For this reason, the next largest asteroid (142)~Polana is used as a central body in order to define a new  family. The HCM at the cut-off velocity of $45ms^{-1}$ listed 11,522 asteroids as members of the Polana family. This group seems to be even more complex, and potentially contains more than a single collisional family. Other works refer to this part of the Hertha clan as the Eulalia family \citep{walsh2013} or even three separate families \citep{dykhuis2015}. 

Distributions of $C$ and $S$-type interlopers within the Hertha clan is shown in Fig.~\ref{img:hertha}. The interlopers of the two spectral types are concentrated in two separate groups, revealing the presence of at least two collisional families. In Fig.~\ref{img:hertha} it could be seen that $C$-type interlopers are mostly located at smaller proper eccentricities, while the $S$-type interlopers groups at somewhat larger values.\footnote{It should be noted that although we are calling these objects interlopers, they are likely members of one of these two overlapping families. Hence, the term interlopers is used to denote that objects for which we know that they do not belong to the $C$ or $S$-type family.} 

In conclusion, we successfully distinguished two separate groups within the Hertha clan, namely the Hertha and Polana families. A further distinction among additional overlapping sub-families requires detailed study of these groups, and probably a somewhat different methodology. This is beyond the scope of our automatic approach, that is designed to be as simple as possible.

\subsubsection{The potential Levin family}

The existence of an asteroid family around (2076)~Levin was first proposed
by \citet{milani2014}. The Levin family is located in the inner part of
the main belt, where both large number density of asteroids and complex
dynamical environment make any attempt to identify families more difficult.
Moreover, it seems that some families share this part of the orbital elements
space, by overlapping each other.

Applying our step one, we found that the concentration of asteroids around
(2076)~Levin merges with background population at $55$~ms$^{-1}$. Thus, our
selection criteria yields to cut-off value of $45$~ms$^{-1}$. It is important
to note here that the Levin family defined at $45$~ms$^{-1}$, also includes
the (298)~Baptistina family, although the latter was treated as a separate
group in \citet{milani2014}.

Actually, the membership of the Baptistina family, as defined at PDS\footnote{http://sbn.psi.edu/pds/resource/nesvornyfam.html} by \citet{nes2015},
exactly matches the membership of our Levin family obtained at $45$~ms$^{-1}$ m/s. A similar definition
of the Baptistina family was also proposed by \citet{masiero2012}. In this respect, there 
is some disagreement whether the Levin and Baptistina are different
families, or they belong to a single family. Therefore, it would be interesting to
see how removal of potential interlopers would influence the situation.

\begin{figure}
	\includegraphics [width=0.48\textwidth]{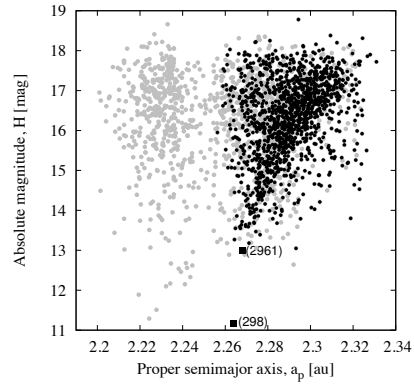}
	\caption{ The distribution of the Baptistina (grey dots) and Levin (black dots) 
	family members in the $a - H$ plane. A one-sided V-shape of the Levin group seems
	to be the most dense region of the outer part of the Baptistina family V-shape.
	This may suggest a common origin of the two groups. Note that asteroid 
	(2961)~Katsurahama is the largest member of the Levin dynamical group.
	}
	\label{f:levin}
\end{figure}

Our search for interlopers among members of the Levin family, performed within stpdf
two and three, did not result in two separated families at $45$~ms$^{-1}$. This might be an indication
that the Levin and Baptistina belong to a single collisional family. However, the large number density 
of asteroids in this region also complicates a reliable identification of interlopers, thus caution is needed.
Still, a close inspection of the family V-shape caused by the Yarkovsky effect suggests
that both groups may belong to the same family (see Figure~\ref{f:levin}).

It is interesting to note that a part of the V-shape is also visible on the left side, for
absolute magnitudes above about $15.5$, in agreement with results by \citet{ivana2015}
who found that most of the objects in this magnitude range should be able to cross the
7/2 resonance with Jupiter.

In summary, the Levin and Baptistina may belong to the same collisional family.
Still, as described by \citet{milani2014}, it might be challenging to explain the current
shape of this group in the proper elements space starting from fragments produced
in a single collisional event. It should be noted however, that this complexity depends
on the cut-off distance used to identified the family, and could be avoided
by adopting different, typically slightly larger, cut-off distances. Therefore,
although there are some reasons to believe that Levin and Baptistina belong to the 
same collisional family, this question remains open. In this respect our approach
of interloper removal does not help much.

\subsection{Other families}
\label{ss:other_fam}

In this Section we briefly discussed some of the results obtained for other families, 
focusing on the definition of the most appropriate cut-off value, i.e. the so-called 
nominal cut-off (see Table~\ref{t:results_final}).

\begin{figure*}
 \includegraphics [width=160mm]{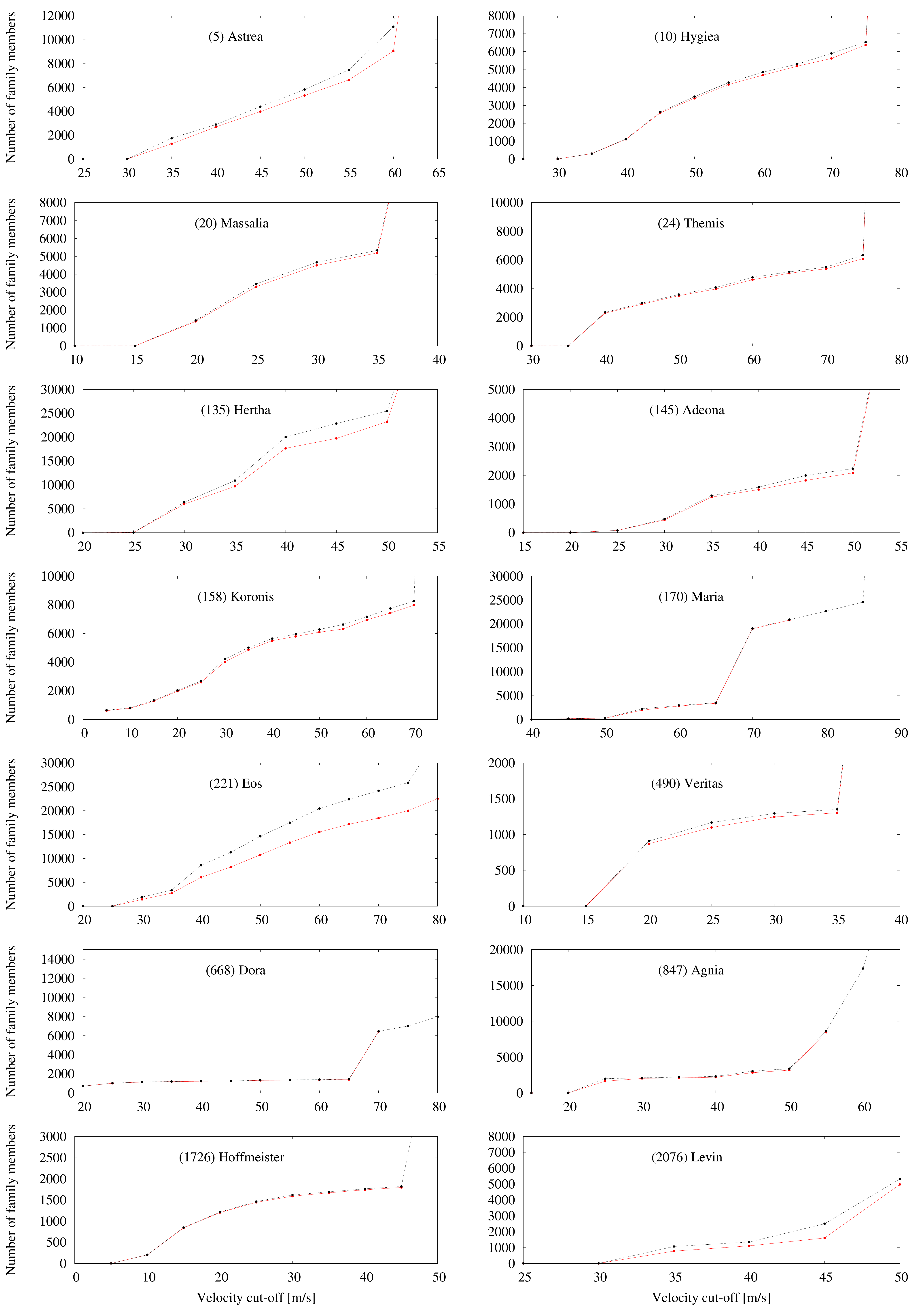}
\caption{Number of asteroids associated to selected families using initial (dashed line) and modified (final) catalogue (solid line). }
\label{f:hcm_all}
\end{figure*}

In most cases the nominal cut-off value is determined using the standard approach, 
i.e. the centre of a plateau (see Figure~\ref{f:hcm_all}). For some families the methodology was somewhat modified, as discussed below.

\begin{table}
 \centering
 \begin{minipage}{90mm}
   \centering
  \caption{Nominal results for selected families. The columns are: (1) Family name (Family), (2) Nominal cut-off value ($d_{nom}$), (3) Number of asteroids initially associated to a family at nominal cut-off (\# initial), (4) Final number of asteroids associated to a family at nominal cut-off (\# final), (5) Percentage of removed asteroids (\%)  }
    \label{t:results_final}
  \begin{tabular}{@{}c|cccc@{}}
  
  \hline
  Family & $d_{nom}$ & \# initial & \# final & \%  \\
 \hline
 (5) Astraea 		& 45 & 4390 	& 3915  & 10.8 	\\
 (10) Hygiea 		& 55 & 4272 	& 4234  & 0.8	\\
 (15) Eunomia 		& 55 & 10042 	& 8128  & 19.1	\\
 (20) Massalia 		& 30 & 4663 	& 4648  & 0.3	\\
 (24) Themis 		& 55 & 4076 	& 3996  & 1.9 	\\
 (135) Hertha		& 35 & 10901	& 9813  & 10.0	\\
 (145) Adeona 		& 45 & 1994 	& 1854 	& 7.0	\\
 (158) Koronis 		& 40 & 5644 	& 5613 	& 0.5\\ 
 (170) Maria 		& 55 & 2229 	& 2146 	& 3.7 	\\
 (221) Eos 			& 35 & 3352 	& 3272 	& 2.4 	\\
 (490) Veritas 		& 25 & 1168 	& 1156 	& 1.0 	\\
 (668) Dora 		& 40 & 1243 	& 1234 	& 0.7 	\\
 (847) Agnia 		& 30 & 2125 	& 2110 	& 0.7	\\
 (1040) Klumpkea 	& 60 & 1739 	& 1568 	& 9.8 	\\
 (1272) Gefion      &  50 & 2541   & 2306 &  9.2 \\
 (1726) Hoffmeister	& 35 & 1692 	& 1686 	& 0.4	\\
 (2076) Levin		& 45 & 2500 	& 2346	& 6.2		\\
\hline
\end{tabular}
\end{minipage}
\end{table}

\textit{(20) Massalia}: The number of asteroids associated to this family is steadily increasing
with cut-off distance, until $40~ms^{-1}$ when the family merges with a local background population,
and the number of associated objects jumps dramatically (see Figure~\ref{f:hcm_all}). Therefore, the nominal cut-off should be between $20$ and $35~ms^{-1}$. We adopt a cut-off of $30~ms^{-1}$ to define
the Massalia family, mainly because the V-shape is well defined in this case. The number of identified interlopers is consistent with a number of expected interlopers estimated by \citet{migliorini1995}. 

\textit{(145) Adeona}: Similarly as in the case of the Massalia family, there is no clear plateau
for the Adeona family in Figure~\ref{f:hcm_all}, although the number of associated asteroids grows
slowly from $35$ to $50~ms^{-1}$. The absence of a clear plateau for this family may be due to the close encounters with massive asteroids that significantly affected the dynamical
evolution of the Adeona family \citep{carruba2003}.

At $55~ms^{-1}$ the number of linked asteroids increases by factor
of a few, and this is because the Adeona merges with the larger Eunomia family as discussed above.
At this cut-off the family also extends on the outer side of the nearby 8/3 resonance with Jupiter,
located at $a=2.705AU$. Examining the V-shape of Adeona family it seems that some family members
should indeed be found on this side of the 8/3 resonance. However, due to proximity of the Eunomia family the nominal cut-off should be selected below $55~ms^{-1}$, and consequently possible
members at $a>2.705AU$ are lost. We adopt the membership found at $45~ms^{-1}$ as the nominal definition
of the family.

\textit{(221) Eos}: This family is unique from many aspects. The albedos of family members cover a range of values somewhere in between the typical values of $C$- and $S$-type, with an average albedo of $\overline{p_v} = 0.13 \pm 0.04$. The obtained $\overline{p_v}$ falls in the ambiguity interval (see Section~\ref{sss:albedo}). Hence, the albedo confidence interval for the Eos family needs to be set manually, and we adopt the $0.1-0.25$ range.

\begin{figure}
	\includegraphics [width=0.44\textwidth]{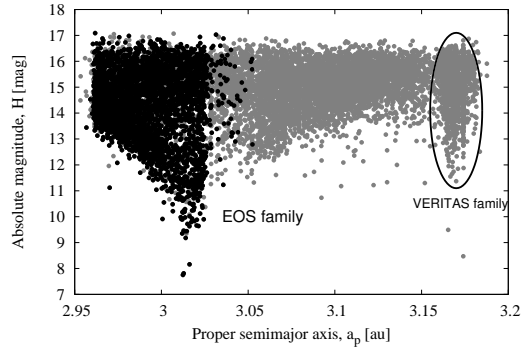}
	\caption{ The distribution of the Eos family members in the $a - H$ plane.
	Black points represent family as identified at $35~ms^{-1}$, while grey points show
	members added at $40~ms^{-1}$.}
	\label{f:eos}
\end{figure}

In the space of proper orbital elements the Eos family is located very close to the Veritas family\footnote{According to \citet{michel2011} asteroid (490)~Veritas might be an interloper in a family named after it. Therefore, the largest member of this family is (1086)~Nata \citep[see also][]{carruba2017}.}, 
and this complicates determination of its membership. These two families merge at 
cut-off of $40 ms^{-1}$, meaning that lower distance should be used to
separate the two groups. However, at $35~ms^{-1}$ almost a half of the Eos family is missing.
This is illustrated in Figure~\ref{f:eos}.
Therefore, a kind of artificial distinction between the Eos and Veritas family is necessary in order
to reasonably define the membership of both of the groups. Still, it should be noted that a smaller
fraction of the Eos family, with $H>15$~mag and $a>3.156$~au, overlaps the Veritas family, suggesting
that the latter may be contaminated by some members of the former group. Also, as noted by \citet{george2017}, 
members of the Eos family could be found at $a<2.95$~au, but a too high
cut-off is needed to jump across the 7/3 resonance with Jupiter that limited the family at the inner edge.
  
\section{Asteroid Families Portal}
\label{s:portal}

The large size and complexity of data acquired in the recent years create a
demand for the new tools necessary to analyse this unprecedented flow of data.
This is because it is often difficult to gather, manage and analyse all these
data. One of the ways scientists are using to cope with these new challenges is the development of
open access web-portals. Such portals, typically devoted to specific purposes, allow
scientists to easily review and analyse large amount of available data.

\begin{figure}
\includegraphics [width=80mm]{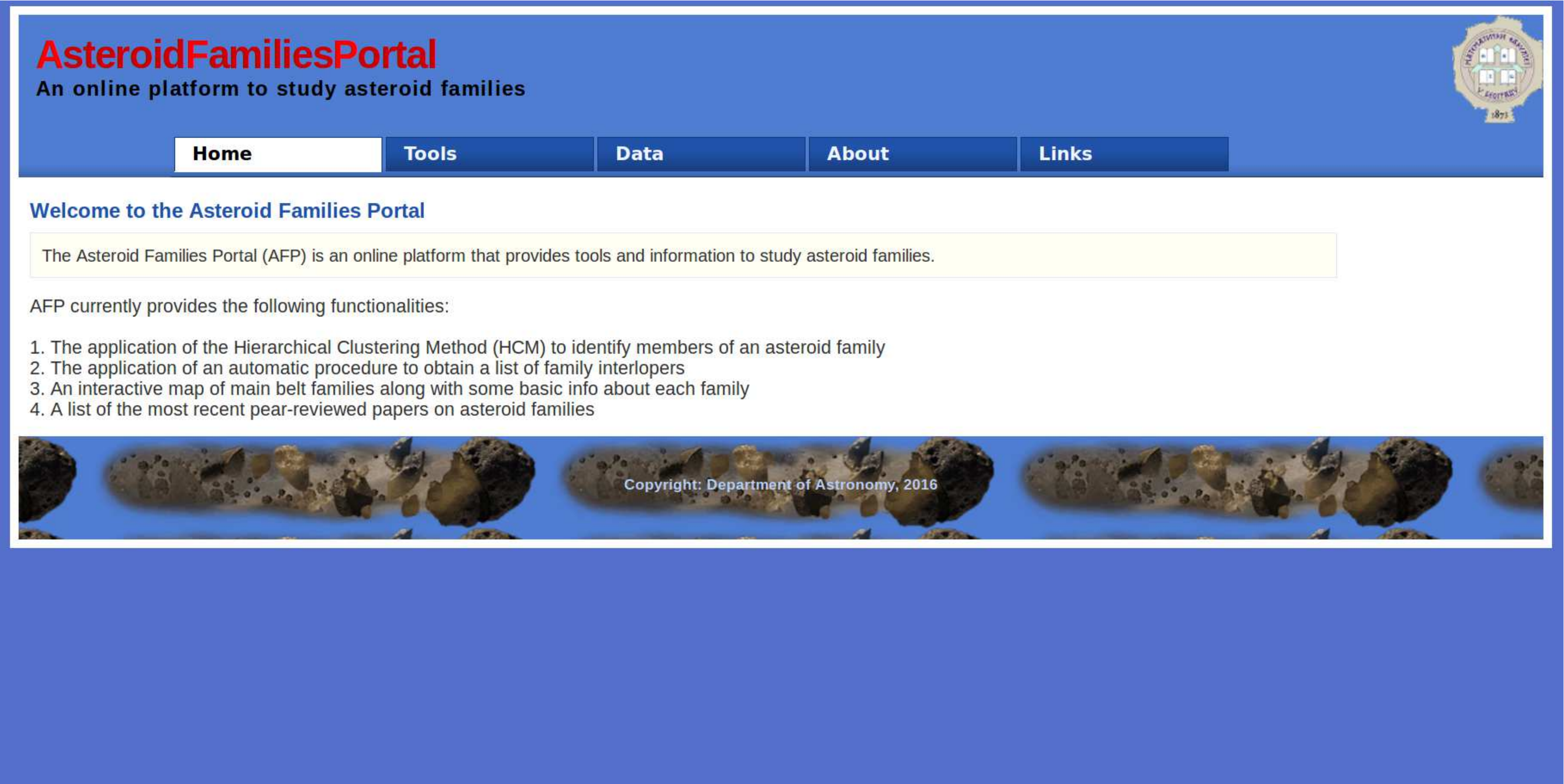}
 \caption{The home page of the Asteroid Families Portal. }
 \label{f:portal}
\end{figure}

In this work we developed and launched this kind of a portal devoted to asteroid families.
The portal is called Asteroid Families Portal (AFP) and is available at this link: asteroids.matf.bg.ac.rs/fam/  
The screen-shot of the AFP home page is shown in Fig.~\ref{f:portal}.

The aim of the AFP is to collect different data about asteroid families and make these freely
available to all interested researchers around the world. It can be used to quickly assess and visualized data. 
Still, this should not be the only
purpose of this portal, but it should also allow to apply different tools and methods
to the provided data. For instance, the well-known Hierarchical Clustering Method (HCM) could
be employed on-line to obtained the most recent list of a family members. Similarly,
the method presented in this work could be used to produce list of potential interlopers
among family members. 

A general idea of the AFP is to have two levels of the functionality for all the available tools, 
the automatic and the advance mode.
The automatic mode is developed to produce the results using the most widely adopted settings
and standards, and is mainly devoted to scientist which are not specialist in families. 
The advanced mode, on the other hand, offers to adjust many parameters used in the computations,
and is primarily devoted to asteroid families specialist.

The AFP is foreseen to be continuously updated and upgraded.
The current datasets will be updated frequently by adding newly available data.
Among the next tools, we foresee for instance to offer a fully automatic estimation
of family age, based on the so-called V-shape method.

We also aim to develop an algorithm that will provide different basic information about
an asteroid family, such is the size of the parent body and its corresponding escape velocity,
the slope of the magnitude-frequency distribution, etc.

\section{Conclusions}

In this work we present an automatic approach to exclude interlopers
from the asteroid families. This approach combines the available data about
spectral reflectance characteristics of asteroids and the iterative application
of the Hierarchical Clustering Method. This algorithm shows very promising 
characteristics, and some advantages with respect to previously used techniques.
The two most important improvements with respect to previous methods are the reduction of the chaining effect,
the well known draw-back of the HCM, and the fully automatic application of our method,
freely available on-line at the Asteroid Families Portal.

There are different possibilities to improve the presented method.
Basically two directions could be followed: i) to refine the way that we are using 
the existing data or ii) to include new data. Regarding the first direction, an option
definitely would be to try discriminate between different spectral types, not only
between the $C$- and $S$-complex. It is known that SDSS data, as used here, does
not allow to distinguish between the $C$- and $X$-type, while with the albedo data
we cannot separate the $S$- from the $X$-type. However, taken together these information
would allow to distinguish between the $C$-, $S$- and $X$-type. Another option would
be to use SDSS based classification proposed by \citet{DeMeoCarry2013}, that makes possible
even finer distinction between the asteroid spectral types.

There are also other data that may be useful to identify interlopers among the asteroid
families. These for instance include the taxonomy based on the new magnitude system as 
proposed by \citep{Oszkiewicz2012}, or polarimetric data that was recently shown to be
very useful to improve taxonomic classification \citep{belskaya2017}.

Finally, our method should benefit a lot from the forthcoming GAIA mission data \citep{wl2007,mignard2007}. 
According to \citet{delbo2012}, for all asteroids observed 
during the GAIA mission, it should be possible to perform multicolour photometry of 
quality good enough for a robust spectral classification.

\section*{Acknowledgements}

The authors would like to thank Alberto Celino, the referee, for his valuable comments 
which helped to improve the manuscript.
This work has been supported by the Ministry of Education, Science and Technological Development 
of the Republic of Serbia, under the Project 176011.
VC acknowledges the support of the S\~{a}o Paulo State Science Foundation 
  (FAPESP, grant 16/04476-8) and of the Brazilian National Research Council
  (CNPq, grant 305453/2011-4).

\label{lastpage}

\end{document}